\documentclass[conference]{IEEEtran}
\usepackage[dvipsnames]{xcolor}
\usepackage[colorlinks=true,linkcolor=blue,breaklinks=True,citecolor=brown,urlcolor=blue]{hyperref}

\usepackage{dingbat}
\usepackage{adjustbox}
\usepackage[linesnumbered,algo2e,ruled]{algorithm2e}
\usepackage{multirow}
\usepackage{booktabs,subcaption,dcolumn}
\usepackage{tikz}
\usepackage{enumitem}
\usepackage{tabularx}
\usepackage[normalem]{ulem}

\usepackage{pifont}% http://ctan.org/pkg/pifont
\usepackage{svg}
\usepackage{soul}
\usepackage[font=footnotesize]{caption}
\usepackage[noadjust]{cite}

\usepackage{amsmath}
\usepackage{amsfonts, amssymb}
\usepackage{colortbl}
\usepackage{tablefootnote}
\usepackage[most]{tcolorbox}
\newcommand{\cmark}{{\ding{51}}}
\newcommand{\xmark}{{\ding{55}}}

\newcolumntype{?}{!{\vrule width 1.5pt}}

\newcommand{\qt}[1]{{\small T#1}}
\newcommand{\qg}[1]{{\small G#1}}

\newtcolorbox{cooltextbox}[1][]{%
    colback=black!5,
    colframe=black!5,
    notitle,
    sharp corners,
    borderline west={0pt}{0pt}{red!80!black},
    enhanced,
    breakable,
    left=0pt,
    right=0pt,
    top=0pt,
    bottom=0pt
    }
    
\newtcolorbox{position}[1][]{%
    colback=black!5,
    colframe=black!5,
    notitle,
    sharp corners,
    borderline west={0pt}{0pt}{red!80!black},
    enhanced,
    breakable,
    left=0pt,
    right=0pt,
    top=0pt,
    bottom=0pt
    }

\newcommand{\textbox}[2]{
    \noindent\fbox{%
        \parbox{0.99\columnwidth}{%
            \textbf{#1}: {#2}
        }%
    }
}

\newcommand{\element}[1]{\textsc{#1}}
\newcommand{\tabelement}[1]{{\scriptsize #1}}

\newcommand\smamath[1]{{\small $#1$}}

\newcommand\smacal[1]{{\small $\mathcal{#1}$}}

\newcommand\smabb[1]{{\small $\mathbb{#1}$}}

\newcommand\mlstab[1]{{\scriptsize \textsc{#1}\xspace}}
\newcommand\mls[1]{{\small \textsc{#1}\xspace}}

\newcommand*\halfcirc[1][0.7ex]{%
  \begin{tikzpicture}
  \draw[fill] (0,0)-- (90:#1) arc (90:270:#1) -- cycle ;
  \draw (0,0) circle (#1);
  \end{tikzpicture}}

\DeclareUnicodeCharacter{0306}{ }

\newtheorem{definition}{\textsc{Def.}}

\newtheorem{remark}{\textsc{Remark}}

\AtBeginDocument{%
  \providecommand\BibTeX{{%
    \normalfont B\kern-0.5em{\scshape i\kern-0.25em b}\kern-0.8em\TeX}}}

\begin{document}

\title{``Real Attackers Don’t Compute Gradients'': Bridging the Gap Between Adversarial ML Research and Practice}

% \titlenote{Produces the permission block, and copyright information}

\author{

\IEEEauthorblockN{{Giovanni Apruzzese\IEEEauthorrefmark{1}, Hyrum S. Anderson\IEEEauthorrefmark{4}, Savino Dambra\IEEEauthorrefmark{5}, David Freeman\IEEEauthorrefmark{2}, Fabio Pierazzi\IEEEauthorrefmark{6}, Kevin Roundy\IEEEauthorrefmark{5}}\\}
\IEEEauthorblockA{{ 
\IEEEauthorrefmark{1}\textit{University of Liechtenstein},
\IEEEauthorrefmark{4}\textit{Robust Intelligence},
\IEEEauthorrefmark{5}\textit{Norton Research Group},
\IEEEauthorrefmark{2}\textit{Meta},
\IEEEauthorrefmark{6}\textit{King's College London}}
\\
{\small \{name.surname\}@\{uni.li\IEEEauthorrefmark{1}, 
nortonlifelock.com\IEEEauthorrefmark{5}, kcl.ac.uk\IEEEauthorrefmark{6}\},
dfreeman@meta.com\IEEEauthorrefmark{2},
hyrum@robustintelligence.com\IEEEauthorrefmark{4}
}}}

\pagestyle{plain}
\maketitle

\begin{abstract}
Recent years have seen a proliferation of research on \textit{adversarial machine learning}.
Numerous papers demonstrate powerful algorithmic attacks against a wide variety of machine learning (ML) models, and numerous other papers propose defenses that can withstand most attacks. 
However, abundant real-world evidence suggests that actual attackers use simple tactics to subvert ML-driven systems, and as a result security practitioners have not prioritized adversarial ML defenses.

Motivated by the apparent gap between researchers and practitioners, this position paper aims to bridge the two domains. 
We first present three real-world case studies from which we can glean practical insights unknown or neglected in research.
Next we analyze all adversarial ML papers recently published in top security conferences, highlighting positive trends and blind spots. 
Finally, we state positions on precise and cost-driven threat modeling, collaboration between industry and academia, and reproducible research. 
We believe that our positions, if adopted, will increase the real-world impact of future endeavours in adver-sarial ML, bringing both researchers and practitioners closer to their shared goal of improving the security of ML systems.
\end{abstract}


%% Main Body
\section{Introduction}
\label{sec:introduction}

Several recent surveys indicate that protecting machine learning (ML) models is not a leading security concern for practitioners~\cite{sun2021mind,boenisch2021never,kumar2020adversarial,bieringer2022industrial,grosse2022so}. 
According to the few recorded accounts of security failures ``in the wild,'' ML systems can be broken by na\"ive attackers that are not systematically exploiting the vulnerabilities of ML, but rather are developing attacks by guessing---either indiscriminately or by some coarse heuristic~\cite{MITREATLAS,tidjon2022threat}.
Red-team exercises on ML systems often take advantage of security gaps that are agnostic to the existence of an ML model, and subsequent defensive recommendations are likewise more broad than, e.g., \textit{adversarial training} \cite{anderson2021practical, mcgregor2021preventing}.  Additionally, the ML models deployed in production-grade ML systems are often not directly observable (and are sometimes even unreachable) by most attackers~\cite{apruzzese2022wild}.

On the other hand, researchers assert that real ML implementations should not follow the principle of ``security by obscurity''~\cite{saltzer1975protection}, and that security evaluations of ML models should assume \text{worst-case} scenarios~\cite{tramer2019adversarial,carlini2019evaluating,arp2022dos}. Indeed, no one can exclude the possibility of future powerful adversaries turning their attention to the ML models embedded in real systems. Fueled by such ``what ifs,'' thousands of papers~\cite{AdversarialML:URL} have showcased successful security violations of ML models by means of sophisticated strategies such as gradient-based algorithms~\cite{Carlini:Towards}.

These few anecdotes suggest that \textbf{the field of ML security suffers from a mismatch between the priorities of practitioners and the focus of researchers.} 
In this position paper we argue that the gap between adversarial ML research and practice is real and identify several underlying causes. We aim to reduce this gap by proposing guidelines for future endeavours that reflect our observations of aspects overlooked by, or inconsistent within, existing literature. 

To reach our goal, we first revisit the fundamentals of ML security (§\ref{sec:revisiting}). We argue that real deployed ML models are part of complex ML \textit{systems} whose architectures are unknown to researchers, and that cybersecurity is rooted in \textit{economic} considerations for both attackers and defenders.

To assist researchers in understanding crucial facets of operational ML security, we then present three original case studies from the real-world (§\ref{sec:casestudies}) that elucidate: 
(a)~the \textit{architecture} of a complex ML system deployed in an online social network; 
(b)~the \textit{lack of evidence} of adversarial examples against a commercial ML system for phishing webpage detection; and
(c)~the role of \textit{time and domain expertise} in devising sophisticated attacks during an acclaimed ML evasion competition.

Next we turn our attention to the research domain and take a snapshot of the current landscape of adversarial ML as portrayed in scientific papers (§\ref{sec:sota}). After surveying the proceedings of the ``Top-4'' security conferences from 2019 to 2021, we {systematically analyze all 88 papers} that consider attacks against ML or corresponding defenses. Of these papers, 89\% only evaluate algorithms based on neural networks, 63\% focus on computer vision, and 80\% perform their experiments on ``benchmarks''. We discover several {inconsistencies} in the terminology adopted in reputable prior work. We also identify several positive trends, such as an increasing amount of papers envisioning attackers who ``ignore'' the ML model and reach their goal by targeting other elements of the ML system.

Finally, we coalesce all our observations by stating four positions that researchers and practitioners in ML security can adopt to close the gap between these domains~(§\ref{sec:recommendations}). We encourage the community to (a)~\text{adapt} threat models to ML systems, (b)~integrate threat models with \text{cost-driven assessments}, (c)~build \text{collaborations} between industry and academia, and (d)~embrace a \text{``just culture''} for source-code disclosure.

In summary, we make three major \text{contributions}: 
\begin{itemize}
    \item We present three \textit{real-world case studies}, showing practical insights unknown or neglected in research.
    \item We \text{analyze all recent adversarial ML papers} in top security venues, highlighting positive trends and blind spots.
    \item We state \text{four positions} that, if adopted, will help bridge the gap between research and practice in ML security.
\end{itemize}
\noindent
Our contributions also include extended discussions (Appendix~\ref{app:extra_cs}) and the detailed findings of our literature review (Appendix~\ref{app:sota}). Our resources can be found in our website~\cite{radcg:website}.
\section{Revisiting Machine Learning Security}
\label{sec:revisiting}

We begin by reviewing the foundations of the three pillars of our paper: machine learning systems (§\ref{ssec:mls}), security of machine learning (§\ref{ssec:secml}), and practical cybersecurity (§\ref{ssec:economics}). We also compare our paper with related work (§\ref{ssec:related}).

\subsection{Overview of Machine Learning (Systems)}
\label{ssec:mls}

\textbf{Machine learning in one paragraph.}
The fundamental goal of machine learning (ML) is to create ``machines that can make decisions by learning from data''~\cite{Jordan:Machine}. At its core, the process entails applying a \textit{learning algorithm} to \textit{training data}, which yields a \textit{machine learning model}. The purpose of any ML model is to \text{generalize} to new data: given an \textit{input}, the ML model produces an \textit{output} that can be used to accomplish a given task, such as predicting future trends, detecting malware, recognizing speech patterns, or inferring objects in images~\cite{Jordan:Machine, Lecun:Deep}. The effectiveness of an ML model is measured during a \textit{validation phase}, during which the ML model analyzes test data and its predictions aer compared to some known \textit{ground truth} (e.g., labels~\cite{joyce2021framework}). Effectiveness can be measured through various metrics (e.g., \textit{accuracy}).
Informally, an ML model belongs to either of two ``paradigms'' that characterize the underlying learning algorithm. Those based on neural networks~\cite{Lecun:Deep} are denoted as \textit{deep learning} (DL), whereas others are broadly denoted as \textit{shallow learning} and include, e.g., SVMs and tree-based algorithms~\cite{Apruzzese:Deep}. 

\begin{figure}[!htbp]
    \centering
    \includegraphics[width=1\columnwidth]{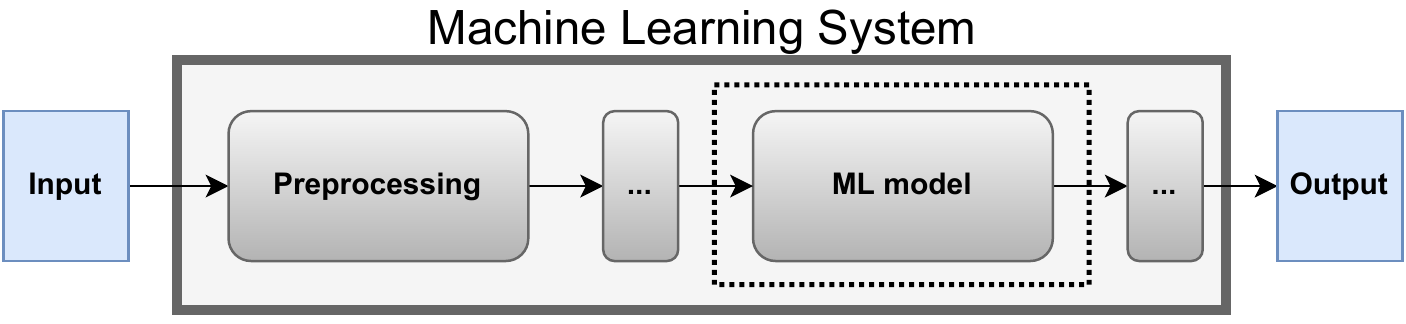}
    \caption{An ML system. The system receives an input, which is preprocessed and then fed to an ML model, the results of which may be further processed before providing the system's final output.}
    \label{fig:mls}
    \vspace{-1em}
\end{figure}

\textbf{Using ML in practice.}
If an ML model is determined to be effective, it can be integrated into a \textit{machine learning system}, which itself can be a component of some product or service. 
Indeed, an ML model by itself is usually insufficient for operational deployment. First, it must receive an \textit{input} that may itself be derived from a separate component (e.g., signal capture or preprocessing\footnote{Preprocessing steps can be complex. They can entail, e.g., transformations, filtering, normalization, feature extraction, and/or noise reduction.}).  Second, the ML model's \textit{output} may be further analyzed for decision making. We illustrate an ML system with a linear data-processing pipeline in Fig.~\ref{fig:mls}.

\vspace{-1em}

Real ML systems may include pipelines having various components, each potentially having its own applications of ML (e.g., image scaling~\cite{xiao2019seeing}, word embeddings~\cite{song2020information}).
Some ML systems require constant updates, which can be done in an automated fashion~\cite{zanella2020analyzing}, while others may adopt company-specific guidelines~\cite{nassi2020phantom}. 
It is even possible that a given input never reaches \textit{any} ML model within the system. This outcome is typical, e.g., in Network Intrusion Detection Systems (NIDS), wherein samples matching known ``signatures'' are put into an alternate pipeline that does not include any ML~\cite{Gardiner:Malware}.

\begin{cooltextbox}
\element{Observation:} A machine learning \textbf{model} is merely a single component within a machine learning \textbf{system}.
\end{cooltextbox}

\textbf{Types of ML systems.}
Many IT infrastructures already rely on ML systems at either small or large scales.
Without loss of generality we identify two main categories of ML systems, characterized by the relationship (i.e., knowledge and degree of interaction) between the ML system and its users.

\begin{itemize}
     \item \textsc{Open.} The source code of these ML systems is publicly available. This category may include custom-built ML systems where the code is subsequently released. Obviously, any ML system is \mls{open} for its developers.
    
     \item \textsc{Closed.} These ML systems are developed in a proprietary setting and distributed to their end users, who can inspect neither (i)~the underlying source code, nor (ii)~the components enclosed in the ML system. These ML systems are either \textit{unrestricted}, if they can be freely used; or~\textit{restricted}, if their use is subject to some restriction, such as a fee or a threshold based on queries. 
\end{itemize}
We provide some examples of these ML systems in Table~\ref{tab:types}.

\begin{table}[htbp]
\centering
    \caption{Types of ML systems with example products and research papers.} 
    \label{tab:types}
    % \vspace{-3pt}
    \resizebox{0.8\columnwidth}{!}{
        \begin{tabular}{c?c|c|c}
            \toprule
            \multirow{2.5}{*}{\begin{tabular}{c}
                 \textbf{Type of} \\ \textbf{ML system}
            \end{tabular}} & \multicolumn{3}{c}{\textbf{Example Application}} \\ \cmidrule{2-4}
            
            & \textit{Product / Company} & \textit{Use case} & \textit{Related Work} \\
            \midrule
            
            \multirow{2}{*}{\mls{Open}} 
            & OpenPilot~\cite{openpilot} & Autonomous Driving & Sato~\cite{sato2021dirty} \\
            & GPT2~\cite{gpt2} & Natural Language &  Carlini~\cite{carlini2021extracting} \\
            
            \midrule
            
            \multirow{2}{*}{\mls{Closed}} 
            & Google Translate~\cite{googletranslate} & Translation (\textit{unrestricted}) & Pajola~\cite{pajola2021fall} \\
            & ClarifAI~\cite{clarifai} & MLaaS (\textit{restricted}) &  Yu~\cite{yu2020cloudleak}\\

            \bottomrule
        \end{tabular}
    }
    
    % \vspace{-3pt}
\end{table}

\begin{figure}[!htbp]
    \centering
    \includegraphics[width=1\columnwidth]{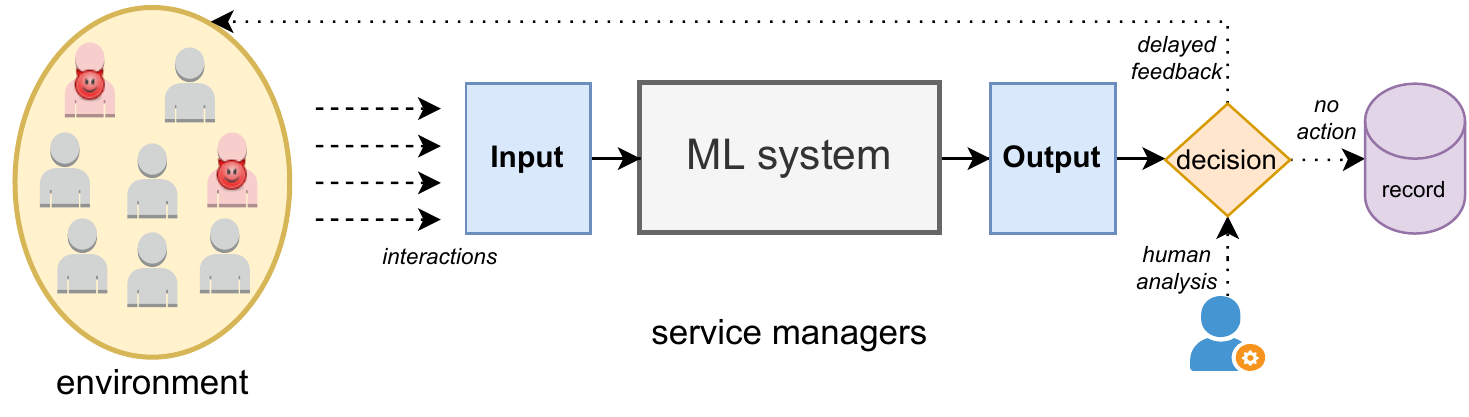}
    \caption{An \mls{{\scriptsize invisible}} ML system. The low privileged users in the \textit{environment} interact with the ML system, the output of which can be used in diverse ways by its \textit{managers}. For example, the alarms of a NIDS may be inspected by sysadmins and then used to block suspicious hosts; whereas a network management system can dynamically allocate available resources (e.g., bandwidth). In some cases the output does not trigger any immediate action. 
    }
    \label{fig:invisible}
    \vspace{-1em}
\end{figure}

Both \mls{open} and \mls{closed} ML systems resemble the schematic shown in Fig.~\ref{fig:mls}: they receive an \textit{input} and provide an \textit{output}. Typically, the users of an ML system can \text{control} the input and can observe the output (e.g., via an API~\cite{nasr2021adversary}). Such properties, however, may not hold for some special cases, which we denote as \mls{invisible} ML systems. 
Control and configuration of \mls{invisible} ML systems are reserved for ``{high-privilege}'' users (e.g., developers or sys-admins), for whom the ML system can be either \mls{open} or \mls{closed}. The functionality of these systems depends on the interactions of ``{low-privilege}'' users (e.g., employees or customers), to whom the operation of the ML system is not readily apparent. A low-privilege user may receive some feedback based on the ML system's output, but such feedback may be influenced by other systems or may arrive after a long and/or unpredictable delay. Simply put, low-privilege users of \mls{invisible} ML systems may not know (i)~\textbf{if} something happened, (ii)~\textbf{when} it happened, (iii)~\textbf{why} it happened, and/or (iv)~\textbf{what} changed. We provide a schematic of an  \mls{invisible} ML system for network management in Fig.~\ref{fig:invisible} (two of our case studies in §\ref{sec:casestudies} entail \mls{invisible} ML systems).

\vspace{-1em}

\subsection{Security of Machine Learning}
\label{ssec:secml}

\textbf{Adversarial Machine Learning.}
Thousands of papers~\cite{AdversarialML:URL} have addressed the security of machine learning, a research field typically denoted as ``adversarial machine learning''~\cite{Huang:Adversarial}. 
Although work in this field has existed for nearly 20 years~\cite{Dalvi:Adversarial}, adversarial ML became popular when Szegedy et al~\cite{szegedy2013intriguing} showed that deep learning is vulnerable to \textit{adversarial examples}, which are examples that are ``similar'' to ``normal'' inputs but induce a given ML model to behave incorrectly~\cite{Biggio:Wild}. 

Adversarial examples can certainly be used in malicious ways: for instance, an attacker can craft an adversarial example that \textit{evades} an ML model powering a security system, e.g., a malware classifier~\cite{Biggio:Evasion}. (We provide the definition of evasion attacks in Appendix~\ref{sapp:evasion}.)
Nevertheless, we will show that adversarial examples are only one of a myriad of threats that can violate the security of ML systems.

\textbf{Threat Model.}
Security assessments require the definition of a \textit{threat model}, which establishes the relationship between an attacker and their target~\cite{apruzzese2022wild}. 
The seminal paper by Huang et al.~\cite{Huang:Adversarial} was among the first to formalize the potential threats to ML. Since then, more efforts followed---some more general (e.g.,~\cite{Biggio:Wild, Papernot:SoK}), others more specific (e.g.,~\cite{apruzzese2021modeling, Laskov:Practical}). In the ML context, threat models characterize an attacker's relationship to an ML model through four elements: \mls{goal}, \mls{knowledge}, \mls{capabilities}, and \mls{strategy}.

\textsc{Goal.} 
    Attackers can have diverse \mls{goals}, such as violating the \textit{integrity} or the \textit{availability} of the ML model~\cite{Demontis:Adversarial}. They can also, however, attempt to ``steal'' the ML model~\cite{tramer2016stealing}, or extract private data from it via \textit{membership inference}~\cite{shokri2017membership} attacks that infer whether a person's data is in the model's training set.
    
\textsc{Knowledge.} 
    Srndiç and Laskov~\cite{Laskov:Practical} pointed out that the attacker's \mls{knowledge} spans over three elements of an ML model: the training data, the feature set, and the algorithm. Depending on the information available to the attacker, most literature adopts a ``box'' terminology~\cite{Papernot:SoK}: ``white-box,'' when the attacker knows \textit{everything}; ``black-box,'' when the attacker knows \textit{nothing}; and ``gray-box,'' for \text{intermediate} cases~\cite{apruzzese2021modeling}.\footnote{We acknowledge that such color-based terminology may offend some people. We use it when discussing prior work to preserve fidelity to the original sources, but we propose that future work refrain from using it (§\ref{sec:recommendations}).}
    
\textsc{Capabilities.}
    Existing literature shows that attackers use various \mls{capabilities} to tamper with ML models. For instance, they can \text{interfere with the training} phase of the ML model~\cite{Biggio:Poisoning}; they can control inputs to a (trained) ML model and \text{observe its output}~\cite{Papernot:Practical}; and/or they can \text{operate from diverse spaces} of the data pipeline of a ML system (the so-called ``problem/input'' and ``feature'' spaces~\cite{Pierazzi:Intriguing}). In some circumstances~\cite{li2020practical}, the attacker cannot interact with the ML model, but they can obtain a ``{surrogate}'' ML model---which is used, e.g., to \textit{transfer} adversarial examples~\cite{Demontis:Adversarial}.
    
\textsc{Strategy.} 
    To reach their \mls{goal}, attackers exploit their \mls{knowledge} and \mls{capabilities} and enact a \mls{strategy}, e.g.: attackers that fully know a deep learning model can use its \text{gradient} to craft adversarial examples~\cite{madry2017towards}, while attackers who can manipulate training data may add a \textit{backdoor}~\cite{yao2019latent}.

We observe, however, that the terminology above has a limitation: it focuses exclusively on the ML \textit{model} (the inner, dotted box in Fig.~\ref{fig:mls}), which is just a single component of the much more complex ML \textit{system} (the outer box in Fig.~\ref{fig:mls}). We will elaborate on how to overcome this limitation in §\ref{sec:recommendations}.

\textbf{Adversarial ML in practice.}
Several researchers have investigated whether the explosive popularity of adversarial ML in research papers has been received with equal interest from industry practitioners. Let us trace the major findings since 2020: % of such findings. % present the evolution of the practitioners' viewpoint with a short story.
\begin{enumerate}[label={\scriptsize ({{\arabic*}})}]
    \setcounter{enumi}{2019}
    \item Kumar et al.~\cite{kumar2020adversarial} conducted a survey of 28 companies. Participants were asked about their perspectives on adversarial ML. Most participants did not know how to respond and were primarily worried about \textit{poisoning} attacks.
    \item The following year, Boenisch et al.~\cite{boenisch2021never} raised an important warning: many ML developers confessed that ``I never thought about securing my ML models.''
    \setcounter{enumi}{2020}
    \item Sun et al.~\cite{sun2021mind} studied ML models deployed on mobile apps and (perhaps unsurprisingly) found that: ``41\% of ML apps do not protect their models at all.''
    \setcounter{enumi}{2021}
    \item A year later, Bieringer et al.~\cite{bieringer2022industrial} interviewed developers of ML and concluded that ``most lack adequate understanding to secure ML systems in production'', even though one third ``feel insecure about adversarial ML.''
    \setcounter{enumi}{2021}
    \item Grosse et al.~\cite{grosse2022so} interviewed hundreds of practitioners asking their opinion on  countermeasures to adversarial ML attacks. The general consensus was ``Why do so?''
\end{enumerate}
We find it surprising that the viewpoint of practitioners has barely changed, even in spite of warnings from sources {beyond} the research domain. For example, in 2020 Gartner predicted that ``by 2022, 30\% of cyberattacks will leverage training-data poisoning, model theft, or adversarial examples.''~\cite{gartner2020trends}

\begin{cooltextbox}
\element{Observation:} Despite abundant evidence showing that ML models are vulnerable, practitioners persist in treating such threats as low priority.
\end{cooltextbox}

\subsection{(Economics of) Cybersecurity}
\label{ssec:economics}

\textbf{An elementary principle.} It is well known that ``there is no such thing as a foolproof system''~\cite{carlini2021poisoning} (the actual quote is from a renowned con artist~\cite{voxFoolproof}). Given a sufficient amount of \text{resources}, any attacker can succeed in their goal~\cite{moore2010economics}. 
Indeed, the purpose of security mechanisms is to \textit{raise the cost sustained by the attacker}~\cite{gordon2002economics}, either by requiring more resources (e.g., time to succeed) or by increasing the consequence of failure (e.g., getting caught). However, \text{countermeasures also have costs} (e.g., implementation efforts, periodic evaluations, maintenance, and computational resources~\cite{anderson2006economics, de2019information}) that must be factored into any decision about whether to build or deploy them. Simply put, operational cybersecurity is rooted in economics~\cite{moore2010economics, wilson2014some}.

\textbf{Cybersecurity in practice.} 
Let us substantiate our previous claims with public statements made by cybersecurity professionals.
On July 22nd, 2022, a tweet~\cite{sophos_twitter} by a Sophos AI employee revealed that ``Adversarial examples are not the primary concern at Sophos.'' This tweet inspired an insightful discussion between renowned researchers and practitioners in this field. Konstantin Berlin, Head of AI at Sophos, explains his thinking about adversarial examples~\cite{sophosKonstantinBerlin} as follows:

\textbox{Head of Sophos AI}{``If you look at cybercrime in economical terms (as you should because it is a business) the optimization for an adversarial ex. is not the expensive part, it is the engineering part of building a tool that can create a diverse set of attacks with no obvious watermarks.'' (\textit{source}:~\cite{sophosKonstantinBerlin_twitter})}

Berlin also stated that ``Given the existence of an already large number of more prevalent attacks that do bypass detections, why prioritize this one?'' 
Such a simple (and recent) use-case shows that operational decisions are dictated by economics, and many practitioners prioritize threats that are deemed to be more important for their businesses.

\begin{cooltextbox}
\element{Observation}: Economics is the main driver of practical cybersecurity---both for attackers and defenders.
\end{cooltextbox}

\noindent
We report the rest of the discussion of such Twitter thread in Appendix~\ref{sapp:twitter}, where we will also provide our own viewpoint.

\subsection{Related Work}
\label{ssec:related}
Many papers share our goal of improving the security of real-world ML systems. The authors of~\cite{arp2022dos} and~\cite{apruzzese2022role} provide practical guidelines on ML \textit{for} security---whereas we focus on the security \textit{of} ML. 
Some related papers (e.g.,~\cite{grosse2022so, boenisch2021never, kumar2020adversarial, bieringer2022industrial}) focus exclusively on the industry perspective; while most literature surveys (e.g.,~\cite{Apruzzese:Addressing, Biggio:Wild, Papernot:SoK}) focus just on the research perspective. We consider both perspectives, because we aim to bridge the existing gap between these two sides. Perhaps the closest effort to our paper is the (unpublished) work by Gilmer et al.~\cite{gilmer2018motivating}, which attempts to contextualize the implications of adversarial ML in the real-world. Despite sharing our goal,~\cite{gilmer2018motivating} has two limitations: first, it mostly focuses on computer vision and deep learning---which, as we will show, represent only a subset of the conceivable applications of ML in reality; second, the remarks made in~\cite{gilmer2018motivating} relate to papers published before 2018---i.e., almost five years ago, when ML deployments were not as popular as today.

We briefly summarize two orthogonal research areas to our paper.
{(i)}~The performance of ML tends to degrade over time~\cite{jordaney2017transcend}-\cite{yang2021cade}. This phenomenon (called ``concept drift'') is due to changes in the data distribution that are physiological in nature and sometimes even beyond an attacker's control. 
{(ii)}~ML can also be used as an attack vector (e.g., ``deepfakes''~\cite{guera2018deepfake}, or ``attribute inference''~\cite{gong2016you}). This research area (i.e., ``\text{offensive ML}'') is outside our scope, because we focus on the protection of ML systems.

To the best of our knowledge, ours is the first position paper (on ML security) whose main \text{arguments} stem from \text{observations} based on (i)~a systematic analysis of recent research, and on (ii)~case studies from real experience.

\section{Case Studies (Real Experience)}
\label{sec:casestudies}

We now describe three case studies fostering the contribution of industry practitioners. Our intent is twofold: (i) to elucidate some practical aspects that may be obscure to researchers, and (ii) to induce practitioners to be more open with their techniques.

The first case study (§\ref{ssec:cs_david}) presents the architecture of a real \mls{invisible} ML system deployed at Facebook, the largest online social network (OSN). 
The second (§\ref{ssec:cs_savino}) describes the triaging of phishing webpages that confused a security company's ML detector.
The third (§\ref{ssec:cs_hyrum}) sheds light on the role of time and domain expertise in devising query-efficient attacks during an ML evasion challenge organized by renowned tech companies.
In Appendix~\ref{sapp:cs_malware}, we provide a fourth case study showing that even ``adversarially aware'' malware detectors can easily be fooled with expert knowledge.

Because our case studies consider ML systems for \textit{cyberthreat detection}, this section will focus on ``{evasion}'' attacks, i.e., misclassifications of malicious input at test time.

\subsection{``The four A's of abuse fighting'':  ML Systems in an OSN}
\label{ssec:cs_david}

This case study reflects the experience of Facebook, a \textit{real} OSN that makes ample use of ML (in various forms) to manage its services and, in particular, to fight spam abuse.

\textbf{Scenario.}
An attacker attempts to spread spam on Facebook; for example, they want to post a pornographic image with some text, which may lure a user to click on an embedded URL. Facebook does not allow these activities on their platform, and hence employs an ML system to prevent this type of content from appearing. The attacker---aware of the existence of the ML system---tries to evade the detector by perturbing the content and/or changing their behavior. 

\textbf{Problem.}
On the surface, this scenario represents the ``evading the ML spam detector'' setting that is typically envisioned in research papers (e.g.~\cite{imam2019survey, rao2021review}). However, most such papers assume that the ML system consists of (i)~a single ML classifier that (ii)~analyzes content and (iii)~outputs the decision to either block or allow the content---a decision which (iv)~is readily observable by the content's author (e.g.,~\cite{madisetty2018neural}). Such a ``black-box'' scenario (which has also been envisioned in its ``white-box'' variant~\cite{grolman2022hateversarial}) assumes an adversary who can send an arbitrary number of requests to the ML model and use its response to craft adversarial examples---potentially with the added constraint of preserving the original input semantics.

\begin{figure}[!htbp]
    \centering
    \includegraphics[width=0.99\columnwidth]{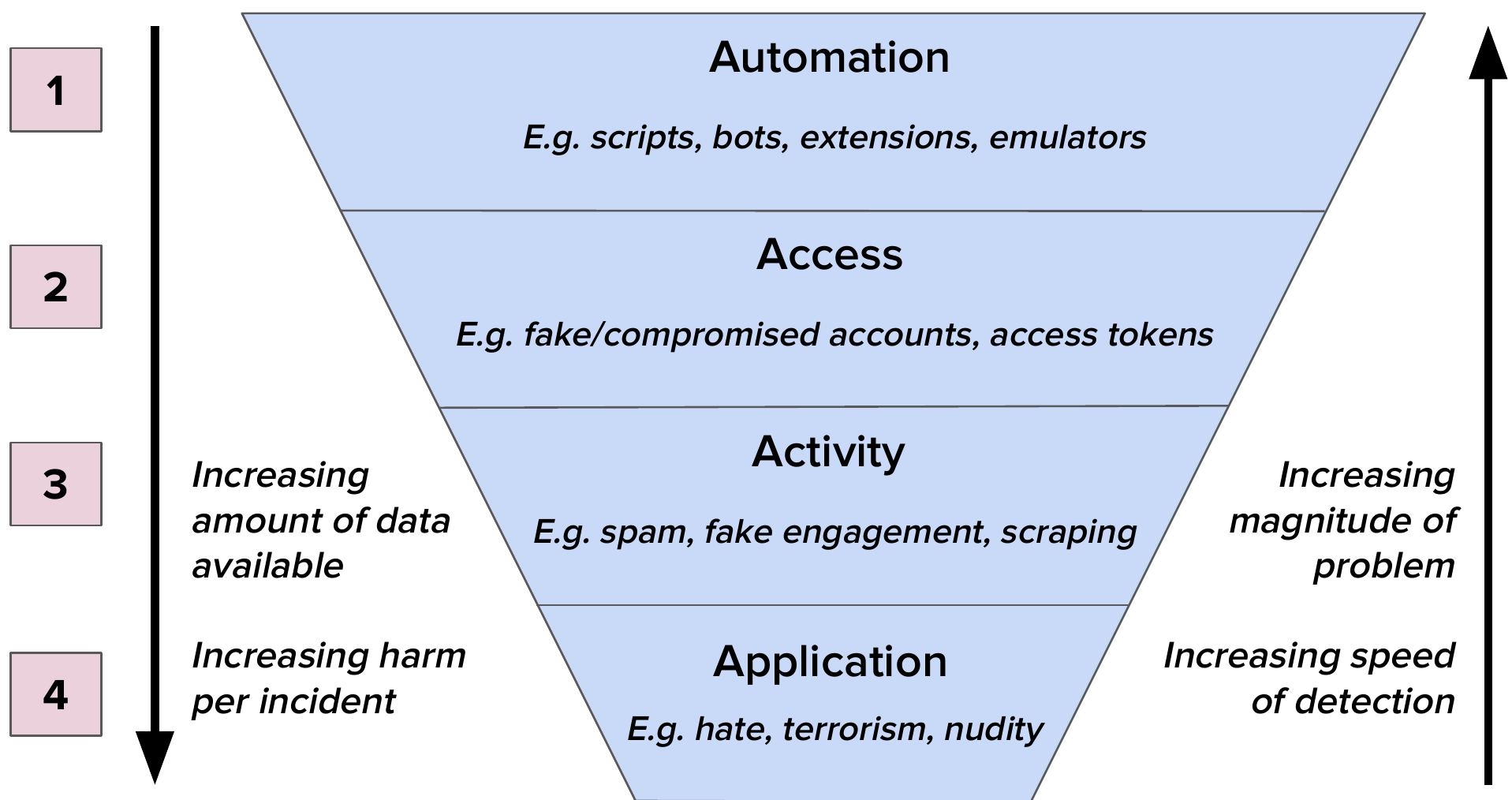}
    \caption{Example of Facebook's ML system for spam detection. The system consists of a ``funnel'' of four interconnected defensive layers, each with its own logic. The attacker must bypass all layers to be successful.}
    \label{fig:meta}
    %\vspace{-1em}
\end{figure}

\textbf{Reality.} 
Production systems however, are far more complex (see §\ref{ssec:mls}). In fact, Facebook views content classification (block/allow) as the \text{last} line of defense in an entire funnel of countermeasures that work in concert to provide defense at different layers; to be successful, an attacker must evade each layer in the funnel.
We denote this funnel as ``the four A's of abuse fighting'': \mls{Automation}, \mls{Access}, \mls{Activity}, and \mls{Application}. We provide an overview in Fig.~\ref{fig:meta} and explain each term below. Without loss of generality we can assume that the input to the ML system is an \textit{action} (e.g., posting an image) executed by a given \textit{entity} (e.g., user account). 

\textit{1) \textsc{Automation} layer.} 
At the top of the funnel is a simple question: ``is the action due to automation?'' Anti-automation is the most general defense because it applies to both read and write surfaces and to logged-in and logged-out traffic. Indeed, Facebook is constantly targeted by bots and automated scripts~\cite{ferrara2016rise}. The \mls{automation} layer automatically prevents ``black-box'' attack strategies based on repeatedly querying the system over short time frames.

\textit{2) \textsc{Access} layer.} 
Automation detection is Facebook's principal defense against logged-out reads at scale. In contrast, an account is required for writes (e.g., {posting} content) or logged-in scraping. Most malicious activity is perpetrated by fake or compromised user accounts~\cite{xiao2015detecting}. 
An attacker can, however, find other ways (e.g., fake advertiser accounts or stolen access tokens) to obtain the required permissions to launch an offense.
Simply put, the role of the \mls{access} layer is to analyze any write request and determine whether the corresponding write permission was obtained legitimately by the requesting entity. 

\textit{3) \textsc{Activity} layer.}
If Facebook fails to stop a malicious action just for being automated (in the \mls{automation} layer) and does not catch the bad actor when they obtain access (in the \mls{access} layer), then the next opportunity is when they begin carrying out their (malicious) activity. %(or, in fact, any activity)
To reach their goal, spammers and other abusers must act in fundamentally different ways from normal users (see Appendix~\ref{ssapp:spammers} for an explanation). Such differences include how they interact with the social graph~\cite{teng_dec}, the distribution of the actions they take, and/or how fast they execute their actions. 

\textit{4) \textsc{Application} layer.} 
The first three layers of the funnel are generic and mostly boil down to well-known cybersecurity practices, some of which may not require ML.
However, if the funnel ended at the third layer, two concerns would arise: (i) What happens if an attacker {bypasses} all three layers? and (ii) What about {specific} prohibited content such as nudity, hate speech, or gun sales---which are activities that, despite being against Facebook's policies, can be produced by ``benign'' users?
To address these concerns, Facebook has a fourth, \mls{application} layer. Assuming that the above layers are mostly effective (i.e., low false positives/false negatives), most of the violations in this layer will be small-scale and driven by real users.  This layer is where the deep learning models typically targeted in research papers~\cite{grolman2022hateversarial} (e.g., nudity detectors or hate-speech classifiers) can really shine.

\textbf{Complexity.} Facebook's spam-detection system has numerous additional complexities not shown in Fig.~\ref{fig:meta}. For example, a blocked attacker typically has no way of knowing which layer ``caught'' them and therefore must try many different evasion strategies. Each layer itself may contain both shallow learning and deep learning methods as well as heuristic rules, further complicating the attacker's task. The layers can inform each other even if an attacker is not blocked---for example, an account may be given a smaller rate limit threshold (\mls{activity} layer) if a fake-account detector (\mls{access} layer) returns a ``suspicious'' score but is not confident enough to block outright. Finally, even if some layers output a decision of ``malicious,'' such output is postprocessed by a \textit{response} layer that decides which action to take: Ban the user? Remove the content? Show a CAPTCHA? Wait and see? The attacker may even observe different responses from multiple instances of the same input, depending on factors beyond the attacker's control---a  characteristic of an \mls{invisible} ML system (§\ref{ssec:mls}). 

\begin{cooltextbox}
\element{Observation:} Real ML systems include many components, not necessarily all using ML. Attackers must bypass all components to be successful.
\end{cooltextbox}

We remark, however, that the complexity of real ML systems does not necessarily mean that such systems are ``omnipotent.'' The following case study will reveal that even full-fledged ML detectors can be thwarted via simple tactics.
\subsection{Are there Adversarial Examples in the Wild (Web)?}
\label{ssec:cs_savino}

To assess the prevalence and nature of adversarial examples ``in the wild'' (i.e., in the real-world), we first consulted the AI Incident Database~\cite{AI_incident_database}, containing over \smamath{1\,600} reports of AI failures (as of Aug. 2022). Unfortunately, querying the database with the keywords ``evasion'' and ``adversarial machine learning'' yielded only \smamath{2} and \smamath{6} results, respectively. Apparently, only a limited number of ({reported}) incidents involve the types of attacks typically portrayed in adversarial ML papers. 
To gain further insight, we asked a leading cybersecurity company whether they encounter adversarial examples when manually triaging security incidents. This case study provides their answer.

\textbf{Context.}
We consider the problem of \textit{phishing webpage detection}. Similarly to the previous case study, the ML system is a commercial-grade detector, composed of diverse modules, all with the underlying goal of ``catching phish.'' Specifically, the ML system (which is \mls{closed} and \mls{invisible} to the attacker\footnote{The attacker can only inspect the {output} by subscribing to the company's security product, fetching the phishing page, and waiting for the back end to recognize the webpage as phishing and eventually add it to a blocklist.}) consists of an ensemble of image classifiers, each responsible for a specific task (e.g., logo attribution, visual comparison, input form detection). The ML system leverages the principle of active learning~\cite{apruzzese2022sok}: For each input, the output is a ``phishing confidence'' score that is used (i) to make an automated dection (i.e., block/allow) and (ii) to improve the ML system by triaging to security analysts inputs for which the ML system is ``uncertain.'' Such a setting fits our objective: most related research focuses on computer vision (§\ref{ssec:sota_overview}) and the ML system naturally suggests to the analysts which inputs can be linked to adversarial examples.

\textbf{Method.}
We searched for adversarial examples in the ML system's usage logs, restricting ourselves to the \smamath{40} domains that (according to the cybersecurity company) were most commonly targeted\footnote{We use the term ``targeted'' to denote a phishing webpage that is crafted so as to resemble a (benign) page of a specific website.} by phishing campaigns in July 2022. During this month the ML system analyzed hundreds of thousands of inputs, and among these \smamath{9\,174} were flagged as ``uncertain'' by the ML system and required manual triage. We used this set as a starting point, and we began to review all such samples (i.e., screenshots). To make our in-depth analysis humanly feasible, as we visually inspected each sample we asked ourselves ``can this be an adversarial example?'' and decided to stop once we obtained a positive answer for \smamath{100} samples. We reached this number after going through \smamath{4\,600} samples (half of our initial population). The discarded samples had nothing in common with adversarial examples, due to being ``out of distribution''~\cite{miller2021accuracy}---i.e., they were significantly different from any sample included in the training data of the ML system (e.g., a domain using a new logo, or shifting backgrounds---typical of, e.g., Netflix). We then used the \smamath{100} positive samples as basis for our in-depth analysis. Our aim was to determine the causes of the ML system's ``uncertain'' response and, in particular, if any cause could be traced back to the gradient-based techniques typical of adversarial examples. This entire process (first inspection and in-depth analysis) was conducted by two authors who worked independently and had regular meetings to resolve issues and arrive at a consensus.

\begin{figure}[!htbp]
    \centering
    \includegraphics[width=0.95\columnwidth]{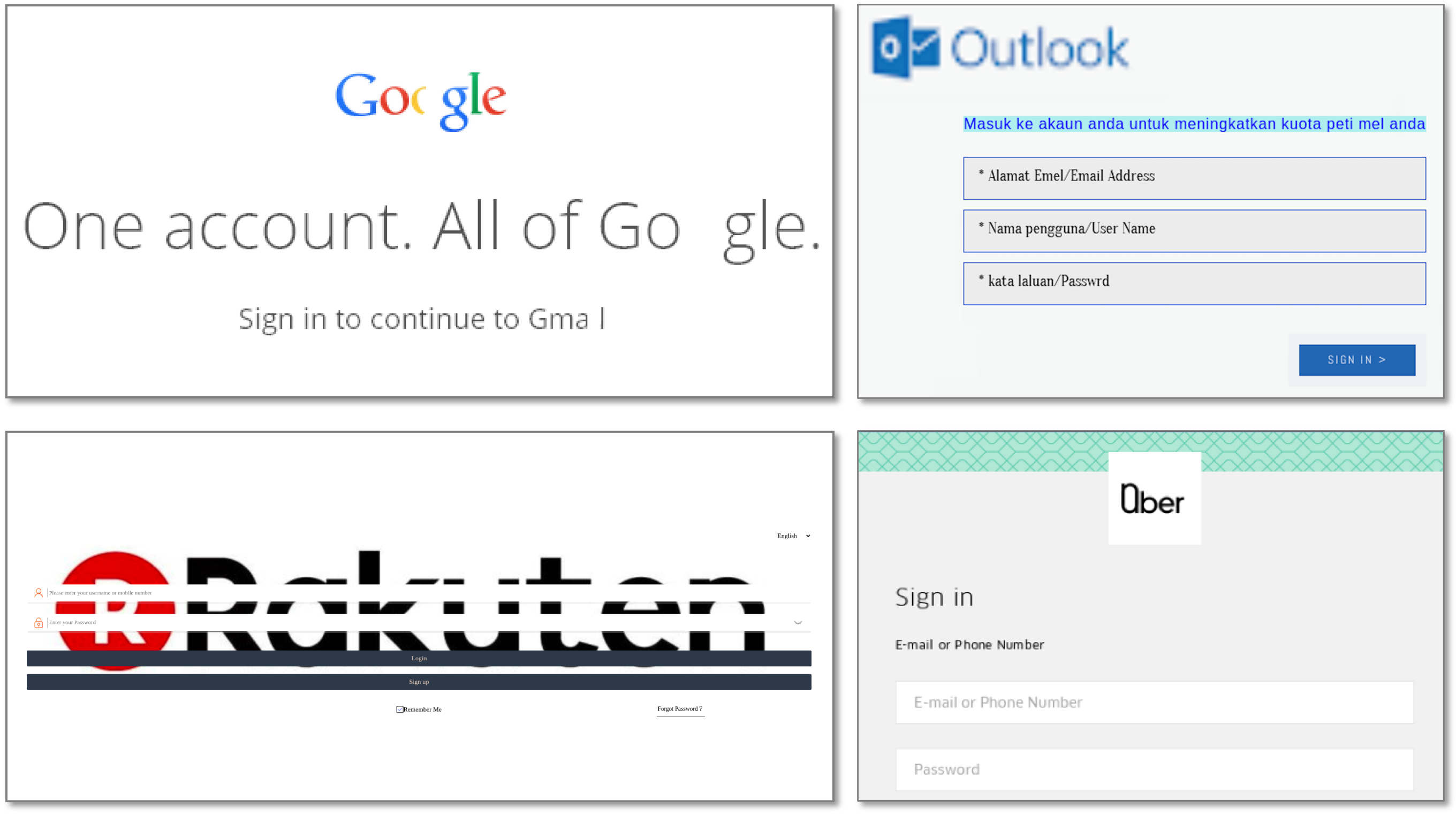}
    \caption{Four evasive phishing samples, depicting the use of masking, cropping, blurring, and misspellings to disrupt the detection of company names, logos, and login-related keywords (e.g., ``Passwrd'').}
    \label{fig:adversarial}
    \vspace{-1em}
\end{figure}

\textbf{Results.}
Our analysis suggests that attackers rely on an arsenal of relatively simple, yet often effective, strategies. We quantify the prevalence of these strategies in Table~\ref{tab:adversarial} and show their effects on real webpages (``in the wild'') in Fig.~\ref{fig:adversarial}.

\begin{table}[!htbp]
    \centering
    \caption{Frequency of different evasive strategies in 100 phishing pages that were poorly analyzed by a commercial ML-driven detector.}
    \label{tab:adversarial}
    \resizebox{0.9\columnwidth}{!}{
        \begin{tabular}{lr|lr}
            \toprule
            \textbf{Evasive Strategy} & \textbf{Count} & \textbf{Evasive Strategy} & \textbf{Count} \\
            \cmidrule(lr){1-2} \cmidrule(lr){3-4}
            Company name style &	25 & Logo stretching & 11\\
            Blurry logo &	23 & Multiple forms - images & 10\\
            Cropping & 20 & Background patterns	& 8\\
            No company name & 16 & ``Log in'' obfuscation & 6\\
            No visual logo & 13 & Masking & 3\\
            Different visual logo & 12 &  & \\
            \bottomrule
        \end{tabular}
    }
    \vspace{-1em}
\end{table}

The most prevalent attacks include cropping, masking, stretching, and/or blurring techniques, all of which induce optical character recognition (OCR) algorithms to incorrectly extract text from the page. Such ``anti-OCR'' techniques have been employed for decades by phishers~\cite{biggio2011survey} and can (still) ably fool ML systems without relying on gradient computations. 
These methods typically target the \textit{company name} when it is part of a logo, mostly by manipulating the logo's visual representation. We also found samples with a different logo altogether, as well as ``company name style'' attacks that alter the logo's company name. Even more rudimentary attacks simply remove the logo and/or the company name (or even both of them in 6 cases).
In rarer cases, we encountered ``multiple form/image'' attacks, which either overlay duplicate forms on top of one another or provide a grid of forms or background images; and ``background pattern'' attacks, which alter the background (especially behind company logos).

\textbf{Considerations.} Our analysis clearly indicates that real adversaries do attempt to evade anti-phishing ML systems that use image classification, and do so with some degree of success. Although our results suggest that the strategies employed by real attackers have little in common with gradient-based adversarial examples, we cannot make such claims with certainty. Indeed, \textit{proving} whether a given ``evasive sample'' is a true adversarial example or just the result of educated guessing is extremely difficult today. This problem could potentially be addressed with digital forensics techniques (which are covered by two recent works in the context of adversarial ML~\cite{shan2022poison, shan2022post}). In a sense, the only way to be 100\% certain that a sample has been generated with a gradient-based strategy would be to ask the attacker directly (i.e., a ``probatio diabolica''). Nonetheless, we are confident in the results of our analysis: the strategies we described above are simple yet effective. Hence, it is sensible to conclude that an attacker would opt for these methods over more computationally expensive ones.

\begin{cooltextbox}
\element{Observation:} to evade phishing ML detectors, attackers employ tactics relying on cheap but effective methods that are unlikely to result from gradient computations.
\end{cooltextbox}

\noindent
As an additional contribution, we release (in our website~\cite{radcg:website}) the \smamath{100} evasive webpages analyzed in this case study.

\subsection{In-Depth Analysis of an ML Evasion Competition}
\label{ssec:cs_hyrum}

We now turn our attention to an ``anti-phishing evasion'' competition organized by industry practitioners in ML security~\cite{cujoevasion}, who provided us with data from the competition. Our intention is to emphasize the role of \textit{time} and \textit{domain expertise} in evading a ML detector. Although most contestants were security enthusiasts, the rules of this competition resembled a constrained and likely scenario simulating a real attack.

\textbf{Rules.} Contestants were given ten phishing webpages (as HTML) and were challenged to manipulate them in such a way as to preserve the original rendering while evading seven ML phishing detectors. These detectors are \mls{closed} ML systems: contestants could input any webpage to the detectors, which provided a ``phishing probability'' (within [0--1]) as output. Winners were determined on two criteria: the number of successful evasions (probability below 0.1) and the number of queries issued to the detectors. Contestants could send an unlimited amount of submissions (i.e., sets of manipulated webpages). A public leader board was regularly updated to reflect the current rankings, showing the number of successful evasions and queries by each contestant. The challenge started on Aug.~6, 2021 and ended on Sept.~17 (42 days in total).

\textbf{Results (high-level).} Four teams crafted variants of all ten phishing webpages that evaded all seven detectors, thereby achieving a perfect score of 70 points. In order of ranking, these teams made \smamath{320}, \smamath{343}, \smamath{608}, and \smamath{9\,982} queries~\cite{cujowinners}.
The 1st-place (two scientists from Kaspersky) and 3rd-place (a single individual) teams published their methodologies~\cite{phishing1st,phishing3rd}. 
We can reasonably assume that the strategy of the 4th-place team involved some automation that resulted in thousands of queries. 
Using query count as a cost metric (as researchers often do, e.g.,~\cite{yu2020cloudleak}), leads to the conclusion that the winning solution had the lowest cost: only \smamath{320} queries, i.e., an improvement of 47\% over the 3rd-place (\smamath{608} queries) and of 97\% over the 4th-place (\smamath{9\,982} queries). 
However, an in-depth analysis shows that this viewpoint is quite misleading.

\begin{figure}[!htbp]
\vspace{-1.5em}
    \centering
    \includegraphics[width=0.95\columnwidth]{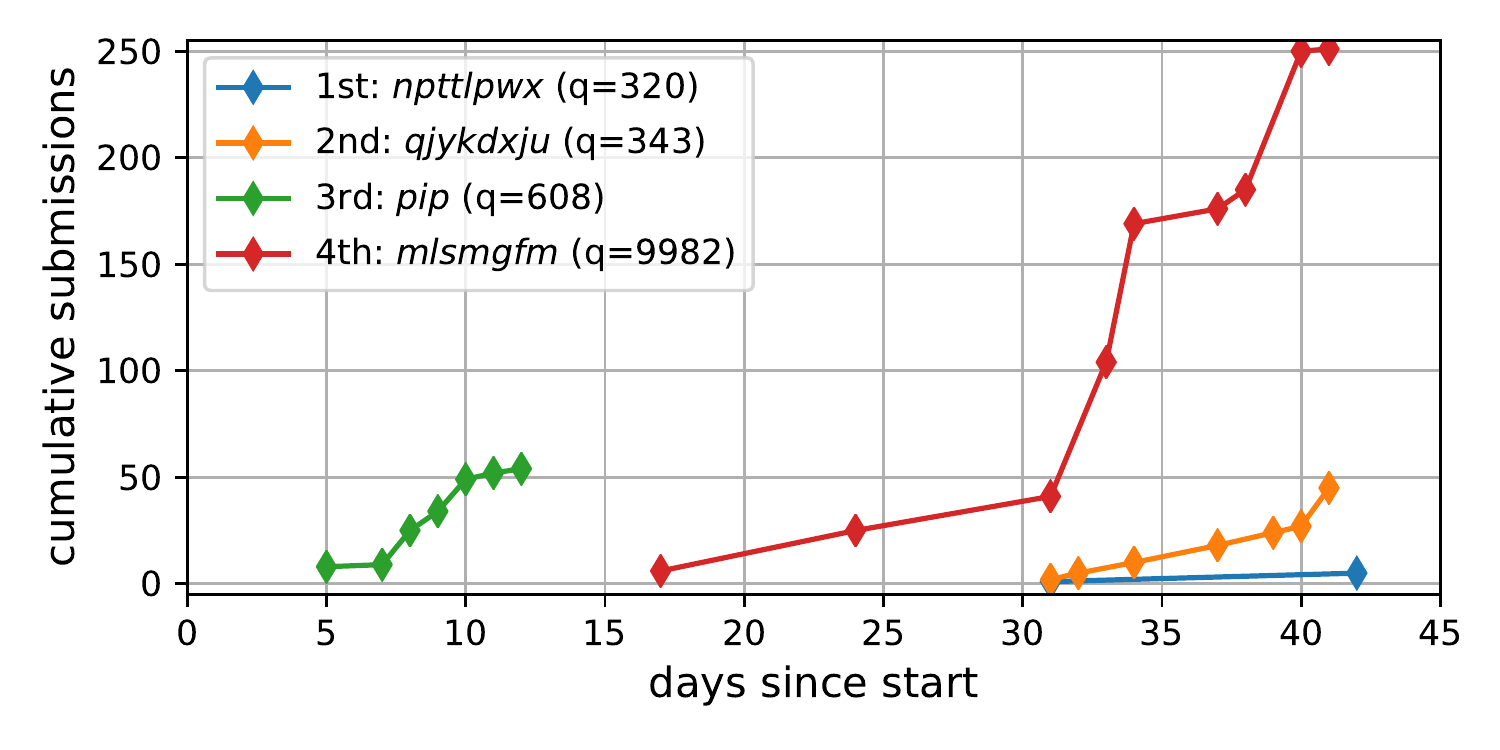}
    \caption{Temporal distribution of cumulative submissions (y-axis) during the phishing MLSEC in 2021 (started on Aug. 6th). Each line indicates a team (q=queries). A detailed explanation of this figure is in Appendix~\ref{sapp:clarification}.}
    \label{fig:phishing}
    \vspace{-1em}
\end{figure}

\textbf{Results (low-level).}
The organizers of this competition provided us with details about the temporal distribution of the submissions made by the four top-ranked teams, whose cumulative submission history is shown in Fig.~\ref{fig:phishing}. We can assume that the \textit{last} submission made by each team corresponds to the point in time in which they ``finished'' their attack.

The 3rd-place team (green line in Fig.~\ref{fig:phishing}) was the first to arrive at a perfect evasion, requiring \smamath{608} queries with the first submission on day 5 and the last on day 12. In contrast, the perfect evasion achieved by the 1st-place team (blue line in Fig.~\ref{fig:phishing}) was the \textit{last} to be submitted---they made their last submission after 42 days, i.e., on the last day of the challenge. Indeed, the winning team reported adopting a ``wait-and-see'' approach, i.e., one that took into account the results of other participants. Originally this team wanted to use typical ``black-box'' techniques based on model replication (e.g.,~\cite{chen2020hopskipjumpattack}), which are query-intensive. However, they {changed their mind}~\cite{phishing1st} when they noticed that the current leader (who ultimately finished 2nd) only had made a few hundred queries. Hence, to win they needed to be very conservative with their API calls. The winning strategy involved sophisticated manipulations of the HTML, rooted in extensive domain expertise. In contrast, the 4th-place team (red line in Fig.~\ref{fig:phishing}) made a large number of queries and submissions, but these were all automated and (likely) required little human effort. Moreover, the 3rd-place team required 12 days for a single person to achieve a perfect evasion, while the 1st-place team required 42 days by two people---an 8x increase in absolute time.

\begin{cooltextbox}
\element{Observation:} Measuring attack efficiency with query counts alone does not reveal the amount of time and domain expertise required to devise a successful offense. 
\end{cooltextbox}
\noindent
We have posted on our website (with permission) the code and anonymized data used in our analysis~\cite{radcg:website}).

%~\cite{hyrumCODE} ~\cite{hyrumDATA}

\textbf{Further Considerations.} So far, we have presented an objective analysis of the evidence available on this competition. However, we believe that there is more to be learned from this case study. We provide further analysis in~Appendix~\ref{sapp:clarification}, and in particular we attempt to derive the human effort entailed in this competition.
Finally, we point the reader to our fourth case study in~Appendix~\ref{sapp:cs_malware}, which further emphasizes the importance of domain expertise on both the offensive and defensive sides of an anti-malware evasion competition.

\section{Snapshot of Adversarial ML Research}
\label{sec:sota}

We now turn our attention to the research domain. We systematically analyze all papers within our scope that have been recently published in selected ``top'' scientific venues. Our objective is twofold: (i)~to identify research \text{trends} and (ii)~to pinpoint \text{blind spots} that may inspire novel studies. 

We first present an overview of our analysis (§\ref{ssec:sota_overview}), and then focus on some positive trends of the threat models envisioned in prior work (§\ref{ssec:tm_attack} for attack papers, and §\ref{ssec:tm_defense} for defense papers); we then highlight some inconsistent terminology adopted in research (§\ref{ssec:tm_confusing}). Appendix~\ref{app:sota} contains a detailed description of our study.

\textbox{Disclaimer}{Our study is not a finger-pointing exercise; it is a holistic review of prior peer-reviewed work aimed at reducing the gap between researchers and practitioners.}

\subsection{Overview}
\label{ssec:sota_overview}

\textbf{Methodology.} We are inspired by Arp et al.~\cite{arp2022dos} and Apruzzese et al.~\cite{apruzzese2022sok}. We consider papers accepted in four venues: ACM Conference on Computer and Communications Security (CCS), USENIX Security Symposium (SEC), Network and Distributed Systems Security Symposium (NDSS), and IEEE Symposium on Security and Privacy (SP). Because we want to focus on \text{recent trends}, we consider the papers published in the last three years: 2019, 2020, 2021. We omit 2022 because (at the time of writing) it is still in progress and we want to see yearly trends.
We identified 88 papers within our scope (shown in Table~\ref{tab:sota}), which we then analyzed.

\textbf{Main Findings.}
We report below the most relevant discoveries, which we will use to support some of the positions taken in this work (§\ref{sec:recommendations}).  Out of 88 papers:
\begin{itemize}
     \item 72\% focus on \text{attacks} (28\% on \text{defenses});
    \item 52\% envision ``{evasion}'' attacks; 
    \item 89\% consider \text{only deep learning} algorithms;
    \item 27\% do not make \text{any} mention of economics;
    \item 51\% publicly release their \text{source code};
    \item 63\% carry out their evaluations on \text{image data} 
      (only 5\% consider malware, phishing, or intrusion detection);
    \item 20\% reproduce a pipeline by building {an ML model only};
    \item 20\% experiment on \text{real} ML systems.
    
\end{itemize}

\textbox{Positive Light}{Taken \textit{individually}, all of these papers are correct. For example, it is legitimate to perform experiments on a self-developed ML system represented by a single ML model that analyzes image data by means of DL. However, some \textit{general} directions taken by the whole body of research generate blind spots---which we attempt to illuminate.}

\noindent
For additional description, interpretation, and figures showing ongoing trends we refer the reader to Appendices~\ref{sapp:questions} to \ref{sapp:specific}.

\subsection{Positive findings in Threat Models of Attack Papers}
\label{ssec:tm_attack}

\textbf{Constrained Adversaries.}
More and more attack papers are considering threat models that incorporate constrained adversaries. For instance, the adversary may have limited knowledge of the targeted ML system, they may have restricted query budgets, or they may be able to observe (or interact with) only a specific segment of the data processing pipeline. The underlying principle is that showing successful attacks in constrained (and thus more realistic) environments is more impactful. In particular, we highlight:
\begin{itemize}
    \item Nasr et al.~\cite{nasr2021defeating} consider attacks against an ML-NIDS. Specifically, they envision a ``blind'' adversary whose perturbations can only be applied on live network traffic, i.e., without knowledge of the packets that will be generated after those manipulated---because such packets will be generated in the (unpredictable) future. This work depicts an exemplary attack against an \mls{invisible} ML system. 
    
    \item Barradas et al.~\cite{barradas2021flowlens} propose FlowLens, an ML system for website fingerprinting. Here, the authors first consider attackers that are oblivious of the existence of FlowLens itself (which are unsuccesful), and then proceed to increase the knowledge/capability of the attacker. Despite it not being surprising that attackers with limited power have little success, we believe it is important to also consider cases of unsuccessful attacks (which is also done in~\cite{carlini2021poisoning}). These evaluations are useful in practice, because they portray attacks that are more likely to occur in the wild due to a lower entry barrier. 
\end{itemize}

We do, however, advocate caution in assuming threat models that envision extremely constrained adversaries. 
Successful attacks stemming from weak adversaries (in terms of \mls{knowledge} and \mls{capabilities}), but requiring sophisticated \mls{strategies}, may not be very realistic---i.e., if the attacker can achieve the same \mls{goal} with simpler \mls{strategies} (see §\ref{ssec:cost_tm}).

\textbf{Unusual Strategies.} 
In the context of evasion attacks, some papers propose \mls{strategies} that have very little in common with established techniques for traditional adversarial examples (e.g., those using gradients~\cite{Papernot:Practical}). We highlight two such papers, both of which consider attacks against ML systems for automated speech recognition (ASR).
\begin{itemize}
    \item Chen et al.~\cite{chen2021real} aim to generate samples that trick both the ML system and humans. To do this, however, they rely on domain expertise, because prior attacks are useless against real ASR. We quote~\cite{chen2021real}: ``However, even with the estimated gradients, none of the existing gradient-based white-box methods can be directly used to attack ASR.'' 
    
    \item Zheng et al.~\cite{zheng2021black} also consider adversaries who (i)~{know nothing} about and (ii)~{cannot query} the target ML system---which is a \textit{real} ASR. The attack is perpetrated through educated guesses stemming from expertise in the audio domain. Indeed, we quote from~\cite{zheng2021black}: ``[because] the ultimate goal of ASR is [...] converting natural speech into text, we believe that the inclusion of the characteristics of natural command audios in the constructed adversarial examples may improve their transferability.''
\end{itemize}
We are pleased to see some research papers that propose evasion strategies that are outside the traditional realm of adversarial ML (which typically focuses on images). 
Indeed, real attackers heavily rely on their domain expertise (§\ref{ssec:cs_savino}).

\textbf{Broader Goals.}
Some papers envision threat models wherein the attack succeeds despite the ML model's correct response. We share two such results targeting production-grade ML systems:
\begin{itemize}
    \item Xiao et al.~\cite{xiao2019seeing} ``evade'' an ML model for object recognition by eliciting an incorrect response of the preprocessing component of the broader ML system. What is intriguing is that the ML model makes the correct prediction (i.e., it outputs the correct label) for the sample received as input---but only because the ground truth of the input sample was changed after preprocessing. This example demonstrates how an attacker can succeed by fooling the ML \textit{system} but not the ML \textit{model}.
    
    \item Nassi et al.~\cite{nassi2020phantom} target the ML-based object detector embedded in autonomous cars. Here, the authors (correctly) guess that such systems are often tuned to prioritize silhouettes of humans (crucial for decision-making in autonomous driving), meaning that these systems can be tricked via ``phantom'' figures with no spatial depth. The authors prove their hypotheses by showing that real autonomous cars stop when a ``phantom'' appears. In this case, \text{the entire ML system makes the correct response}, but the attacks are successful due to specific policies adopted by the system's developers.
    \end{itemize}
Simply put, attackers' goals need not require fooling the ML model or even the entire ML system. We are encouraged to see research that is exploring these scenarios.

\begin{cooltextbox}
\element{Observation:} Assuming constrained adversaries is common. Increasingly more attacks leverage domain expertise. In some cases, attackers' goals go beyond merely causing ML models to misclassify samples.
\end{cooltextbox}

\subsection{Positive Findings in Threat Models of Defense Papers}
\label{ssec:tm_defense}

Only 24 papers (out of 88) consider defenses. Compared with attack papers, defense papers tend to put a stronger focus on the \textit{cost} of the countermeasure, typically by measuring its overhead (i.e., baseline performance degradation).

\textbf{Defending against knowledgeable attackers.}
A common approach in defensive papers is to consider adversaries with perfect knowledge (e.g.,~\cite{hussain2021waveguard,du2021cert}). Such ``white-box'' attacks can be difficult to stage in reality (as also hinted in §\ref{ssec:tm_attack}) because acquiring complete knowledge of the targeted ML system can be expensive. In some cases, however, such knowledge can be easily acquired, especially if the targeted ML system is \mls{open} (see, e.g., the attacks by Sato et al.~\cite{sato2021dirty}). Indeed, creating an \mls{open} ML system is increasingly dangerous, because doing so exposes the system to attacks at different layers.  For instance, Bagdasaryan and Shmatikov~\cite{bagdasaryan2021blind} show how to ``poison'' an ML system by manipulating its source code, i.e., without any change to the training data. 

\textbf{Defenses against limited-knowledge attackers.}
Tang et al.~\cite{tang2021demon} propose a defense that considers (and is evaluated against) attackers who are powerful, but not omniscient; a similar ``black-box'' threat model is adopted also in SIGL~\cite{han2021sigl}.
Some (e.g.,~\cite{arp2022dos, carlini2019evaluating}) may argue that similar defenses violate the ``Kerchoff principle''~\cite{kerckhoffs1883cryptographie}, but we believe that the line of research taken by these works (\cite{tang2021demon,han2021sigl}) is worthy of being pursued due to its high value for real-world deployments of ML. Indeed, attackers without perfect knowledge are a likely threat in the wild, and hence corresponding defenses also demand attention {in research}. 
Notably, some works have shown that attackers with limited knowledge may be as dangerous as omniscient ones: Song and Mittal~\cite{song2021systematic} show that ``the gap between white-box attack accuracy and black-box attack accuracy is much smaller than previous estimates~\cite{nasr2019comprehensive}.''

\begin{cooltextbox}
\element{Observation:} Most defenses assume attackers with perfect knowledge by default. Some, however, are tailored for attackers with limited knowledge.
\end{cooltextbox}

We further discuss and motivate our stance on defenses tailored for limited-knowledge attackers in Appendix~\ref{ssapp:defense}.

\subsection{Inconsistencies that May Harm Future Research}
\label{ssec:tm_confusing}
During our analysis, we focused our attention on the definitions of the threat models envisioned in each paper. We report some of the confusion we encountered, with a focus on the \mls{knowledge} and \mls{capabilities} of the envisioned adversaries.

\textbf{What does the attacker {know}?}
The (very common) terms ``white-box'' and ``black-box'' are used in different works to denote significantly different degrees of attacker \mls{knowledge}. We highlight such inconsistencies by reporting (verbatim, emphasis ours) the descriptions provided in some papers, all of which focus on ``evasion'' attacks (for consistency).

\begin{itemize}
    \item According to Co et al.~\cite{co2019procedural}: ``In white-box settings, the adversary has complete knowledge of the model architecture, parameters, and \textit{training data}.[...] In a black-box setting, the adversary has no knowledge of the target model and no access to \textit{surrogate datasets}.'' This description aligns with the one by Srndic and Laskov~\cite{Laskov:Practical}. 
    
    \item Shan et al.~\cite{shan2020gotta}, however, consider a different ``white-box'' setting: ``We assume a basic white box threat model, where adversaries have direct access to the the ML model, its architecture, and its internal parameter values [...] but \textit{do not have access to the training data}.''
    
    \item Xiao et al.~\cite{xiao2019seeing} do not specify anything about the training data: ``In this paper, we focus on the white-box adversarial attack, which means we need to access the target model (including its structure and parameters).'' As a matter of fact, Xiao et al.~\cite{xiao2019seeing} also consider ``black-box'' attacks, which target an unknown ML model but which is trained on the exact same dataset as the ``white-box'' scenario.  
    
    \item Suya et al.~\cite{suya2020hybrid} assume a ``black-box'' attacker that ``does not have direct access to the target model or knowledge of its parameters,'' but that ``has access to pre-trained local models for the same task as the target model'' which could be ``directly available or produced from \textit{access to similar training data.}'' Such a definition is in stark contrast to the ``black-box'' one by Co et al.~\cite{co2019procedural}.
\end{itemize}
We also report the definition of the ``gray-box'' setting of Hui et al.~\cite{hui2021practical} which ``gives full knowledge to the adversary in terms of the model details. Specifically, except for the training data, the adversary knows almost everything about the model, such as the architecture and the hyper-parameters used for training. Note that the adversary cannot know the training data (called a whitebox).'' This definition aligns with that of Shan et al.~\cite{shan2020gotta}---for a ``white-box'' threat model!

\textbf{What is the ``box''?} 
Whenever ``box''-based terminology is used, it is crucial to establish what is actually denoted by the ``{box}.'' Recall that in the real world, attackers interact with ML {systems} (i.e., the dotted rectangle in Fig.~\ref{fig:mls}) and not with ML {models}. Hence, a ``white-box'' attacker (resp. ``black-box'') should have complete (resp. zero) knowledge of the entire \textit{system}. This consideration is overlooked in research papers: for instance, Zhao et al.~\cite{zhao2019seeing} aim to attack ``real world object detectors'' and propose ``white-box''/``black-box'' attacks---but only consider the perspective of the single ML model. A similar case can be made for the descriptions by Demontis et al.~\cite{Demontis:Adversarial}: here the attackers' goal is first presented in terms of ``\textit{system} operation,'' but ten lines later the knowledge is presented with a ``white-box'' terminology defined as follows: ``the attacker has full knowledge of the target \textit{classifier}.''
Additional confusion stems from the so-called ``no-box'' attacks envisioned by Abdullah et al.~\cite{abdullah2021sok}, whose definition conflicts with the original one by Li et al.~\cite{li2020practical}.
Specifically, the ``no-box'' attacker in Abdullah et al.~\cite{abdullah2021sok} knows \textit{nothing}, which is the de facto assumption of ``black-box'' attackers. In contrast, Li et al.~\cite{li2020practical} shift the focus from the \mls{knowledge} of the attacker to their \mls{capabilities}: they cannot interact with the box (i.e., the ML model), but {know} that it behaves similarly to a surrogate ML model developed by the attacker, who also {knows} a subset of its training data.

\textbf{What can the attacker \textit{do}?}
The attacker's \mls{capabilities} were also difficult to identify. In some cases, this confusion resulted from imprecise  usage of the term ``access.'' For example, Shan et al.~\cite{shan2020gotta} state that their attacker must have ``direct access to the ML model'' but do not specify whether such access includes \textit{read} or \textit{write} privileges; Suya et al.~\cite{suya2020hybrid} state that the (``black-box'') attacker has ``no access to the target model,'' but has ``API access to the target model'' (both statements are in the same sentence). 
Another issue we encountered concerns the format of the \textit{output} received by a given attacker in query-based strategies. Output can come in many forms~\cite{jagielski2020high}, but most papers merely state that the output is ``a probability'' without providing additional details.

\begin{cooltextbox}
\element{Observation:} Definitions of the \mls{knowledge} and \mls{capabilities} of the envisioned attackers are inconsistent, especially when using ``box''-based terminology.
\end{cooltextbox}

\section{Recommendations}
\label{sec:recommendations}

Insofar, we have made a number of observations, stemming from our preliminary discussion (§\ref{sec:revisiting}), our real-world case studies (§\ref{sec:casestudies}), and our detailed analysis of recent research~(§\ref{sec:sota}). We summarize these observations in Table~\ref{tab:observations}. 

\begin{table}[!htbp]
    \centering
    \caption{List of original \textsc{observations} made in our paper.} 
    \label{tab:observations}
        \begin{tabularx}{0.99\columnwidth}{c|X|c}
             \toprule
             \textsc{\#} &~~\textsc{Observation} & Ref.\\
             \midrule

             1 & \tabelement{ML models are only one component of ML systems.} & {\scriptsize §\ref{ssec:mls}} \\ 
             
             2 & \tabelement{Academia and industry perceive adversarial ML differently.} & {\scriptsize §\ref{ssec:secml}} \\ 
             
             3 & \tabelement{Economics is the main driver of practical cybersecurity.} & {\scriptsize §\ref{ssec:economics}} \\

             4 & \tabelement{Evasion is achieved by bypassing all layers of an ML system.} & {\scriptsize §\ref{ssec:cs_david}} \\ 

             5 & \tabelement{Evidence of adversarial examples in the wild is scarce.} & {\scriptsize §\ref{ssec:cs_savino}} \\ 
             
             6 & \tabelement{Queries are not always an effective measure of attack cost.} & {\scriptsize §\ref{ssec:cs_hyrum}} \\  
            
             7 & \tabelement{Attackers use domain expertise and have broad goals.} & {\scriptsize §\ref{ssec:tm_attack}} \\ 
             
             8 & \tabelement{Defenses can envision either strong or weak attackers.} & {\scriptsize §\ref{ssec:tm_defense}} \\ 
             
             9 & \tabelement{Terminology is often imprecise and/or inconsistent.} & {\scriptsize §\ref{ssec:tm_confusing}} \\ 
             
             10 & \tabelement{Evading some ML systems can be very simple.} & {\scriptsize App.\ref{sapp:cs_malware}} \\ 
             \bottomrule
        \end{tabularx}
    %}
\end{table}

We can now link all these contributions and use them to support four actionable positions: {adapting} threat models to ML systems (§\ref{ssec:adapting}), integrating threat models with {cost-driven} assessments (§\ref{ssec:cost_tm}), encouraging industry and academia to {collaborate} (§\ref{ssec:collaboration}), and embracing a ``{just culture}'' for {reproducible} research (§\ref{ssec:disclosure}).

\textbox{Disclaimer}{We phrase all our positions with ``must'' because this is what \textit{we} advocate to bridge the gap between research and practice. However, not embracing {any} of our positions does not invalidate future contributions.}

\subsection{Adapting Threat Models to ML Systems}
\label{ssec:adapting}

Our first position is motivated by the (unintended) inaccuracies shown in some papers (§\ref{ssec:tm_confusing}), as well as by most papers tunnel-visioning on a single ML model (§\ref{ssec:sota_overview}) and overlooking the much more complex ML systems (§\ref{ssec:cs_david}). Note, however, that {complexity} does not imply {robustness} to evasion (as shown in our case study in Appendix~\ref{sapp:cs_malware}).

\begin{position}
\element{Position:} 
Threat models must \textbf{precisely} define the viewpoint of the attacker on \textbf{every component} of the ML system.
\end{position}

\textbf{Holistic vision.}
The elements typically used when describing ML threat models (i.e., \mls{goal}, \mls{knowledge}, \mls{capability}, \mls{strategy}) must be extended to cover the entire ML system. 

\textsc{Goal+.} Real attackers have broader \mls{goals} than just attacking a single ML model. For example, they may want to cause damage to all tenants of an ML system~\cite{apruzzese2022wild}, or simultaneously fool humans and ML classifiers~\cite{mirsky19ctgan,schneider2022concept}.

\textsc{Knowledge+.} To better portray real threats, \mls{knowledge} assumptions should cover all components of ML systems, e.g., any preprocessing steps~\cite{Pierazzi:Intriguing}. For instance, an attacker who has \textit{perfect knowledge} of an ML model but \textit{zero knowledge} of the preprocessing will have \textit{limited knowledge} of the ML system. (We provide our viewpoint on the interplay between \mls{knowledge} and domain expertise in Appendix~\ref{ssapp:domain}.)

\textsc{Capabilities+.} As with \mls{knowledge}, attacker \mls{capabilities} must consider all components of the ML system. For example, attackers should only be assumed to have the ability to give arbitrary inputs to the ML model (e.g., irrespective of preprocessing) if doing so is possible in reality (especially if operating only in the feature space~\cite{apruzzese2021modeling}). 
Moreover, the \textit{output} of the ML system must be explicitly defined (e.g., label-only, label-and-score, top-$k$-scores, all-scores, logits~\cite{jagielski2020high}).
Finally, manipulations of the training dataset should also consider the components of the ML system that collect such data.% in the first place.

\textsc{Strategy+.} By extending their vision to the whole ML system, an attacker can adopt \mls{strategies} that ``ignore'' the specific ML models. For example, Hong et al.~\cite{hong2019terminal} apply manipulations at the hardware layer, Bagdasaryan et al.~\cite{bagdasaryan2021blind} manipulate the source code used to develop the ML model, and Batina et al.~\cite{batina2019csi} derive additional information by spying on electromagnetic side-channels. Notably, Rakin et al.~\cite{rakin2021deep} describe their threat model (in terms of \mls{knowledge} and \mls{capabilities}) both from the system and the {ML} perspectives---a promising first step towards more holistic threat models.

\textbf{Precise terminology.}
As the foundation of every security assessment, a threat model must be unambiguous---especially in research papers. A poorly defined threat model is detrimental because it can lead to wrong estimations of attack power (e.g., the attacker may appear weaker), as well as unfair comparisons by future work. In particular, we identify four terms that are subject to misinterpretation from either a researcher-to-researcher or a researcher-to-practitioner viewpoint. These terms are: \mls{box}, \mls{access}, \mls{evasion}, \mls{adversarial}. 

\textsc{Box}. We have already highlighted inconsistencies revolving around ``box''-based terminology (§\ref{ssec:tm_confusing}). Our stance is that researchers should \text{refrain} from using this terminology.  Instead, we recommend (i) using \textit{perfect}, \textit{limited}, and/or \textit{zero} to define the attacker's \mls{knowledge}, and (ii) precisely specifying the components of the system to which the terms apply.

\textsc{Access}. We have already discussed some issues with this term (§\ref{ssec:tm_confusing}). Our stance is that researchers should (i) be precise when categorizing access (e.g., \textit{read} or \textit{write}), avoiding ambiguous terms such as ``direct'' or ``internal,'' and (ii) specify the type of access for \textit{every} component of the ML system.\footnote{An attacker can, for example, have \textit{query} access to the ML system, meaning that they can only send an input and observe the output of the entire system; or the attacker can have \textit{write} access to only the training data, implying no access to any other component of the ML system.} 

\textsc{Evasion}. In related literature, this term denotes attacks which (i)~cause misclassifications and which (ii) occur ``at test time.'' In contrast, the generic definition of ``evasion'' in cybersecurity is to \textit{bypass a detection mechanism}~\cite{handley2001network}, i.e., to cause a malicious sample to be classified as benign.\footnote{In statistics terminology, an evasion corresponds to a ``type II error.''}
For instance, backdoor attacks~\cite{yao2019latent} are considered to be ``poisoning'' {in the literature}~\cite{Biggio:Wild}, not ``evasion.'' However, from a security standpoint, a (real) attacker does not want to ``poison'' a dataset---rather, they may use ``poisoning'' (e.g., backdoors) to achieve their goal of evading the ML system.
Our stance is that ``evasion'' should be used exclusively to denote the attacker's \mls{goal}, which can be achieved via many strategies (e.g., poisoning~\cite{yao2019latent}, exploiting adversarial examples~\cite{tong2019improving}, or by ignoring the ML model completely~\cite{hong2019terminal}).  

\textsc{Adversarial}. In research, it is typical to link the term ``adversarial'' with the ML context---most likely due to the popularity of ``adversarial examples.'' In many conversations with practitioners, however, we found that this term generates a lot of confusion: in a security context, ``everything is adversarial.'' Our stance is to use this term only when referring to established notions,\footnote{``Adversarial attack'' and ``adversarial threat model'' are tautologies and can be replaced with ``attack against ML'' and ``threat model'' (Appendix~\ref{ssapp:adversarial}).} such as ``adversarial example'' or ``adversarial ML.'' 
Nonetheless, we make an important remark, inspired by~\cite{apruzzese2022role}: \textit{adversarial examples are not malicious per-se}; what is malicious is {crafting and using} an adversarial example. Studying adversarial examples can have uses orthogonal to security, such as improving robustness of ML methods~\cite{fischer2019dl2}.

\subsection{Cost-driven Security Assessments}
\label{ssec:cost_tm}

Our second statement is motivated by the poor consideration given by recent papers (see §\ref{ssec:sota_overview} and Appendix~\ref{sapp:g4}) to the economics of cybersecurity: 27\% do not mention ``cost'' at all.

\begin{position}
\element{Position:} Threat models must include \textbf{cost-driven} assessments for both the attacker and the defender.
\end{position}

\textbf{Context.} 
Biggio and Roli~\cite{Biggio:Wild} point out that the best way to approach cybersecurity is through {proactive defense}: the system designer should first model the adversary, simulate the attack, and then develop a countermeasure ``if the attack has a relevant impact.'' We stress, however, that all of these steps should also account for the cost to both the attacker and the defender. For instance, developing a countermeasure just because an attack has ``high impact'' may not be wise if the corresponding attack is unlikely to happen in the first place. System designers, when conducting their security assessments, prioritize threats that are more likely to occur in the wild (see §\ref{ssec:cs_savino}). Real attackers, as pointed out by Wilson et al.~\cite{wilson2014some}, operate with a cost/benefit mindset: they will only attempt to evade a detector if they perceive the benefits to outweigh the costs. At the same time, {no defense is foolproof}~\cite{carlini2021poisoning}.

\textbox{Disclaimer}{From a \textit{research} perspective, it is appropriate to assume (and evaluate) attackers with perfect knowledge who break any ML system. From a \textit{practical} perspective, however, security evaluations are costly, and it is reasonable that real companies prioritize more common threats~\cite{nistAIrisk}.} 

\begin{figure}[!htbp]
    \centering
    \begin{subfigure}[t]{0.3\columnwidth}
        \centering
        \frame{\includegraphics[width=1\columnwidth]{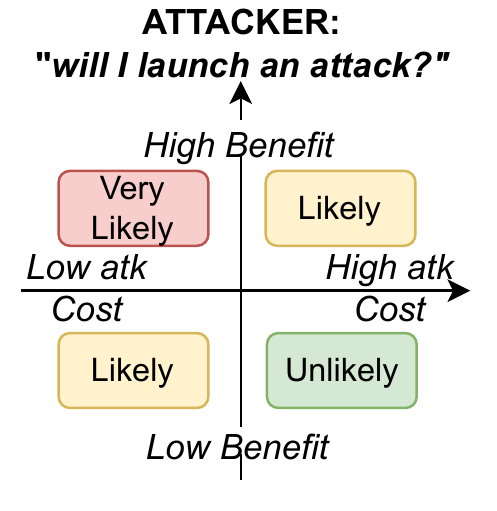}}
        \caption{Attacker perspective: \textit{very likely} threats yield \textit{high benefit} with \textit{low cost}.}
         \label{sfig:attacker}
    \end{subfigure}%
    ~ 
    \begin{subfigure}[t]{0.68\columnwidth}
        \centering
        \frame{\includegraphics[width=1\columnwidth]{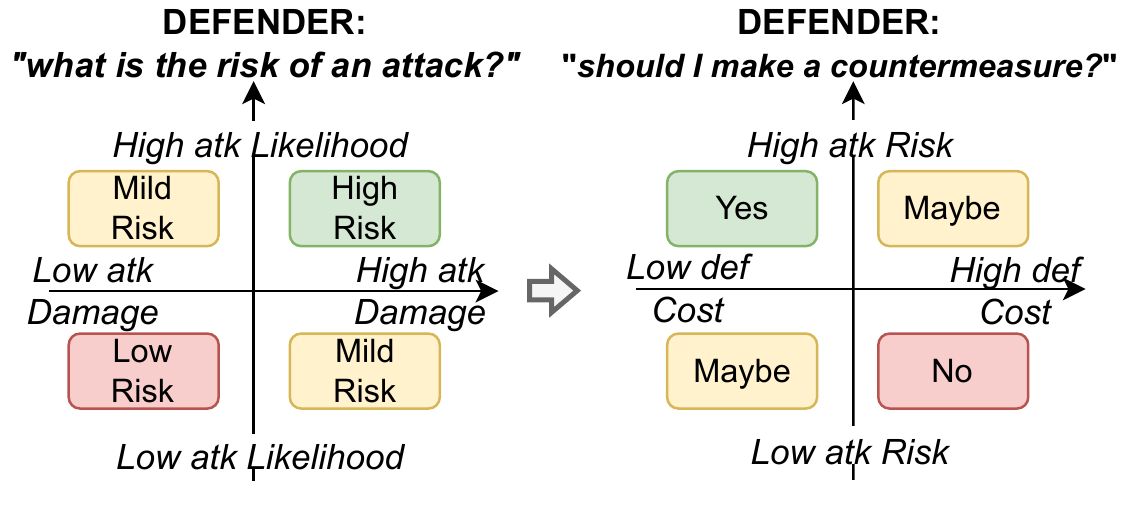}}
        \caption{Defender perspective. The attack's \textit{risk} is assessed via its \textit{likelihood} and its \textit{potential damage}. On the basis of such risk, a \textit{countermeasure} is developed if its \textit{cost} is acceptable.}
         \label{sfig:defender}
    \end{subfigure}
    %\vspace{-3mm}
    \caption{Cost-driven threat modeling. The proactive security lifecycle~\cite{Biggio:Wild} should include economic considerations for both the attacker and the defender.}
    \label{fig:costmodel}
    \vspace{-1em}
\end{figure}

\textbf{Proposal.}
To promote future research that has a greater impact \textit{in practice}, we extend the recommendations of Biggio and Roli~\cite{Biggio:Wild} with respect to \textit{cost-driven} threat modeling. We provide an overview in Fig.~\ref{fig:cost}. When establishing a given threat model, the system designer (i.e., the researcher) weighs the potential {benefit to the attacker} and the corresponding {cost to reap such benefit} (Fig.~\ref{sfig:attacker}).  This preliminary analysis yields the {likelihood} of the attack~\cite{nistAIrisk}. The designer can then simulate the attack and gauge the corresponding {damage}, yielding the \textit{risk}~\cite{lee2021cybersecurity, van2021threat, woods2021sok} of the attack (Fig.~\ref{sfig:defender}, left). Finally, the designer can conceive of any given countermeasure, estimate its {costs} (e.g., implementation, overhead, upkeep), and---depending on the risk of the corresponding attack---assess whether each countermeasure should be deployed or not (Fig.~\ref{sfig:defender}, right). Simply put, quoting from~\cite{apruzzese2022role}, the driver of cybersecurity is ``Paying $x$ (for the defense) to avoid paying $y$ (if the attack succeeds), with $y \gg x$.''

We make three remarks on our cost-driven threat modeling:
\begin{itemize}

    \item Although 57\% of papers attempt to measure the ``cost'' of an attack/defense, they often overlook the human cost factor (e.g., time and expertise (§\ref{ssec:cs_hyrum})). Hence we encourage future studies to \textit{account for human effort}. A possible (albeit imperfect) way of doing so is by computing the time needed to write the source code for a given attack/defense~\cite{robles2006beyond}.\footnote{An intriguing question is: are defenses with low computational overhead \textit{in research} (e.g.,~\cite{shafahi2019adversarial, wong2019fast}) also cheap to deploy \textit{in practice}?} This computation can be facilitated by, e.g., using GIT repository's commit history~\cite{mcdonald2013performance, oliveira2020code} or by process-mining techniques~\cite{caldeira2019assessing}.
    
    \item It is wrong to consider the attacker/defender battle as a zero-sum game in economical terms. Indeed, in many cases, the gain of a (successful) attacker may not correspond to an equal loss by a (broken) defender. We discuss a (toy) use-case in Appendix~\ref{sapp:game} that makes use of the fact that cybersecurity is typically outsourced~\cite{kshetri2021economics}.
    
    \item Researchers proposing ``novel'' attacks should assume the viewpoint of a real adversary. An attack will not be launched if---despite initial success---the attack can be defused via simple heuristics.
\end{itemize}

Finally, a precise and holistic threat model (§\ref{ssec:adapting}) is greatly beneficial for cost-driven assessments \textit{in research}. Future work can use such descriptions to estimate the (cumulative) cost of the attack at a later point in time, and then use such an estimate as reference for comparison (even if the threat model envisions a nation-state adversary---see Appendix~\ref{ssapp:nation}).

\subsection{Collaborations between Industry and Academia}
\label{ssec:collaboration}

Our third position stems from the facts that (i) only 20\% of our analyzed papers (§\ref{ssec:sota_overview}) consider real ML systems, and (ii) 90\% of papers only consider deep learning algorithms, which are not necessarily those used in practice (e.g., owing to a long history of tabular data stored in production databases). 

\begin{position}
\element{Position:} Practitioners and academics must \textbf{actively collaborate} to pursue their common goal of improving the security of ML systems.
\end{position}

The gap between research and practice would diminish if researchers had easier access to production-grade ML systems.
Indeed, most real ML systems are \mls{invisible} or \mls{closed}, preventing researchers from truly understanding how they work (as also mentioned in the Sophos Twitter thread---see \cite{sophos_twitter} and Appendix~\ref{sapp:twitter}). 
For example, a researcher can reasonably assume that an OSN uses an (\mls{invisible}) ML system for spam detection (e.g., §\ref{ssec:cs_david}); however, without knowing its internal architecture, they could hardly develop a ``realistic'' ML system with equivalent functionality. Similar issues exist also for \mls{closed} ML systems. Consider the attack by Pajola et al.~\cite{pajola2021fall}, in which specific character modifications induce Google Translate to output a wrong translation. From the researcher's viewpoint, it is impossible to determine precisely \textit{why} the attack was successful. (E.g., was it due to a bug in the preprocessing phase or in the ML model itself?) 

We acknowledge that asking tech companies to publicly release their own ML systems is unrealistic (although some do, e.g.,~\cite{microsoftopensource}). However, companies can still make it \textit{easier} for academics to do security assessments. For instance, a significant barrier faced by researchers is \textit{getting in contact} with practitioners. 
A potential compromise could be to offer ``bug-bounty'' programs that clearly define guidelines for interaction between researchers and platforms (e.g.,~\cite{facebookwhitehat}), and/or dedicating resources to processing researcher requests (e.g.,~\cite{youtuberesearch}); in both cases companies benefit by gaining thorough, independent security assessments of their ML systems. 
Another possibility would be providing high-level schematics of ML systems (as in §\ref{ssec:cs_david}), inspiring researchers to run experiments on more realistic ML pipelines.

\subsection{Just Culture and Reproducible Research}
\label{ssec:disclosure}

Our last position stems from the observation that only half of the papers we analyzed release their source code (§\ref{ssec:sota_overview}).  
We acknowledge that there may be legitimate reasons to avoid complete disclosure (see Appendix~\ref{sapp:unreproducible}), but we conjecture that this low number is in part due to fear of being criticized.

\begin{position}
\element{Position:} 
In ML security, source-code disclosure must be promoted with a \textbf{just culture}~\cite{dekker2009just}.
\end{position}

A ``just culture''~\cite{dekker2009just} denotes an approach in which mistakes are assumed to occur and derive from organizational issues. Mistakes are avoided by understanding their root causes and using them as constructive learning experiences. This definition is in contrast to a \textit{blame culture}~\cite{khatri2009blame}, which seeks to avoid mistakes by actively blaming the perpetrators, with the intention of discouraging individuals from making mistakes out of fear of reputational damage. We believe that our community should avoid the latter and embrace the former.

A just culture encourages source-code disclosure, enabling the gradual improvement that is the foundation of research. In the ML security context, some attacks (or defenses) may be incorrectly implemented or evaluated, as was the case with DeepSec~\cite{ling2019deepsec}, whose evaluation was found to be flawed \textit{after} its publication~\cite{carlini2019critique}. Some of these flaws, however, were spotted \textit{because} the implementation of~\cite{ling2019deepsec} was publicly accessible. Fear of criticism could have induced its authors~\cite{ling2019deepsec} to not disclose the details of their platform---but \textit{they rightly released their code}, and our community learned from such mistakes, \textbf{turning a negative result into a positive outcome}. (We are glad that~\cite{ling2019deepsec} is still included in SP19's proceedings.) 

Although we endorse future researchers to apply as much rigor as possible when performing their evaluations (i.e., before submitting their papers), we believe that authors should not be afraid of publicly releasing their source code due to the risk of others discovering flaws. Indeed, \textit{there is much to learn from flawed implementations}: the ML domain constantly evolves, and errors can be systematized into practical guidelines (e.g.,~\cite{pintor2021indicators}) to help future research avoid mistakes. 

\section{Conclusions}
\label{sec:conclusions}

As our positions are adopted, we hope to see (i) more reproducible research that (ii) describes threat models for entire ML systems using precise terminology, (iii) makes cost-driven assessments factoring in human effort, and (iv) fosters active collaboration with practitioners from industry. As a final remark, we also endorse the development of novel techniques for the forensics of ML incidents: perhaps real attackers \textit{do} compute gradients---but we cannot prove it yet!

\textbf{\mls{Acknowledgements.}}
The authors thank all participants of the Dagstuhl Seminar ``Security of Machine Learning''~\cite{dagstuhl}, as most of the positions described in this paper derive from discussions originated during this event. The authors also thank the anonymous reviewers for their insightful feedback and valuable discussions, and the Hilti Corporation for funding.

% \clearpage


\appendices

\section{Additional Case Studies and Considerations}
\label{app:extra_cs}

\subsection{Evasion via Adversarial Examples (and Perturbations)}
\label{sapp:evasion}
We provide here the formal {definition} of an evasion attack via adversarial examples and discuss the {perturbations} typically used to craft an adversarial example.

\textbf{Definition.}
Let $M$ be a (trained) ML model taking inputs in some vector space $\mathcal{V}$ and producing predictions in some label set $\mathcal{L}$.  An \textit{adversarial example} is a tuple $(x, \varepsilon) \in \mathcal{V} \times \mathcal{V}$ with the following properties:
\begin{enumerate}
\item $x' = x + \varepsilon$ has the same ground truth label as $x$.
\item $M$ outputs the correct label for $x$ (i.e., if $y \in \mathcal{L}$ is the ground truth label for $x$, then $M(x) = y$).
\item $M$ outputs an incorrect label for $x'$ (i.e., if $y$ is as in (2), then $M(x') \neq y$).
\end{enumerate}
We call $x$ the \textit{original input}, $\varepsilon$ the \textit{adversarial perturbation}, and $x'$ the \textit{adversarial example}.

This notation describes an ``evasion attack'' against ML; i.e., a misclassification at test time.
Given $x$, finding a suitable perturbation can be expressed as an optimization problem (i.e., computing $\varepsilon$ subject to some constraints). 

\textbf{Consideration.}
In related literature, the adversarial perturbation \smamath{\varepsilon} used to craft an adversarial example is typically very ``small'' (i.e., imperceptible by humans~\cite{Papernot:SoK}). However, as pointed out by several works~\cite{carlini2019evaluating, apruzzese2022role}, this constraint does not hold universally. Some attackers may want adversarial examples to be \textit{noticed} by humans~\cite{schneider2022concept}), while in other domains (e.g., malware~\cite{Pierazzi:Intriguing}) humans do not inspect inputs to the ML model, meaning that perturbations can be of higher magnitude---as long as they are physically realizable~\cite{tong2019improving}.

\subsection{Researchers vs Practitioners (Twitter thread in §\ref{ssec:economics})}
\label{sapp:twitter}
We document here the Twitter thread~\cite{sophos_twitter} mentioned in §\ref{ssec:economics} and provide our viewpoint on this discussion.

\textbf{Discussion (verbatim).}
We consider the comments between Battista Biggio (B), a well-known researcher in adversarial ML~\cite{Biggio:Poisoning, Biggio:Wild}; Joshua Saxe (J), chief scientist at Sophos AI; and Konstantin Berlin (K), head of Sophos AI.

\begin{itemize}[label={\small J1)}]
    \item {\small Why robustness to adversarial examples isn't a first-priority concern on the Sophos AI team. [image showing the `simplistic' vision of a ML malware detector by researchers, compared with the `complex' pipeline of a real ML malware detector.]}
\end{itemize}
\begin{itemize}[label={\small B1)}]
\item {\small I’d tend to agree. But still optimizing over a set of transformations in a black-box manner may help find blind spots even in more complex systems -- surely lp norms are not that useful here -- I am talking about combinations of more interesting transformations that abuse the PE.}
\end{itemize}
\begin{itemize}[label={\small K1)}]
    \item {\small Given the existence of an already large number of more prevalent attacks that do bypass detections why prioritize this one?}
\end{itemize}

\begin{itemize}[label={\small B2)}]
    \item {\small The goal of all these defenses in the end is about raising the bar for the attacker, and this work in my opinion is useful in that direction. It is important that people know that ML is not learning what we expect it should learn…}
\end{itemize}

\begin{itemize}[label={\small K2)}]
    \item {\small If you look at cybercrime in economical terms (as you should because it is a business) the optimization for an adversarial ex. is not the expensive part, it is the engineering part of building a tool that can create a diverse set of attacks with no obvious watermarks. [follows] Think of Sophos as human adversarial attackers against your adversarial attack. My bet is that if you deployed most published ML adversarial attacks at scale Sophos would block you within a few hours in such a way that you would have to rewrite your PE generator, no ML will help.}
\end{itemize}

\begin{itemize}[label={\small B3)}]
    \item {\small It depends. The transformations used are perfectly legit (e.g. adding sections). Btw we never thought of bypassing a real AV pipeline, we know that is more complicated. We only wanted to show that some ideas (using DL from raw bytes) are much more insecure and need improvement.}
\end{itemize}
The last message (B3) was sent on July 24, 2022 at 10:06 AM, while the first message (J1) was sent on July 22, 2022 at 8:17 PM (both times are GMT+1).

\textbf{Our viewpoint.} 
This discussion underscores our observation that researchers and practitioners are solving different problems, or at least are working in different threat models. 
The tip-off is one specific term in K2:
``if you deployed most published ML adversarial attacks \textit{at scale}.'' Our first case study (§\ref{ssec:cs_david}) offers an example of threats that operate ``at scale'' being prioritized by companies, and it is not surprising that Sophos would likely do the same. However, the attackers envisioned in many adversarial ML papers (and, more generally, in many security papers~\cite{urbina2016limiting, erba2020constrained, li2019stealthy}) tend to be subtle. Hence, both sides are making valid points, but they do not appear to agree because they are not talking about the same thing.

Despite this gap (which our paper attempts to close), we believe that B3's point is valid. 
Clearly (§\ref{ssec:collaboration}), a better cooperation between industry and academia would kick-start novel research efforts that assess the robustness of full-fledged ML systems against adversarial examples. However, there is still much to learn from the results of recent literature, despite the fact that the envisioned ML systems are often inaccurate representations of real ones.

\subsection{Further Considerations on our third case study (§\ref{ssec:cs_hyrum})}
\label{sapp:clarification}
To allow a more complete understanding of our anti-phishing evasion case study (§\ref{ssec:cs_hyrum}), we provide a further description of the {rules} of this competition. We also provide our interpretation of the events that transpired during the challenge when taking into account the human effort (and decision making) required by each top-ranked team.

\subsubsection{Queries and Submissions}
We clarify Fig.~\ref{fig:phishing} by explaining the difference between a ``submission'' ($y$-axis) and a ``query'' (legend).
During the challenge, participants could either: \textit{query} the detectors---either a single one or all of them (there were 7 detectors in total); or \textit{submit} their (adversarial) webpages. The latter counts as a ``submission'' ($y$-axis) and automatically generates 7 queries for each webpage included in the submission. For example, a participant that submits 2 webpages automatically triggers 14 queries, because each webpage is analyzed by all 7 detectors (to determine how much points to award).
The organizers of MLSEC logged each submission but not the individual queries (i.e., the single API calls). This is why the plot in Fig.~\ref{fig:phishing} shows the submissions in the $y$-axis. We are aware that a more fine-grained analysis would consider the history of the individual queries. However, as is evident by Fig.~\ref{fig:phishing}, there is a strong correlation between number of 
submissions and number of queries (although, technically, one could issue 10k queries and make only one submission), so we reported submissions in the $y$-axis.

\subsubsection{Interpretation (human effort)}  
Let us summarize the \textbf{facts} (for brevity, we denote the teams as 1st, 2nd, 3rd, 4th).

\begin{itemize}
    \item On day 0 all teams had a common objective, to use as few queries as possible.
    \item On day 5, 3rd made his first submission.
    \item On day 12, 3rd made his last submission, thereby completing his attack and leading the ranking with \smamath{608} queries. This information was publicly available: after this moment, all teams teams knew that, to win, they had to use fewer than \smamath{608} queries.
    \item From day 17 to day 42, 4th made various submissions and reached a perfect evasion by making \smamath{9\,982} queries.
    \item On day 31, both 1st and 2nd made their first submission. (This does not mean that they did nothing before day 31.)
    \item On day 41, 2nd made their last submission, producing a perfect evasion with \smamath{343} queries. This information was publicly available.
    \item On day 42, \textit{7 hours after 2nd's last submission}, 1st made their second submission, which included 9 webpages (out of 10), all of which evaded all the ML detectors.
    \item Also on day 42, 1st made their last submission, producing a perfect evasion with \smamath{320} queries.
    \item At the end of the challenge, 1st stated~\cite{phishing1st}: ``At first, we thought of attempting a classic model replication attack [...] but as we started working on the competition, we noted that the leader had already achieved the highest possible score using just \smamath{343} API calls''
\end{itemize}

Now, let us \textbf{interpret} these facts. Firstly, the statement by 1st is technically incorrect, because the first submission by 1st was on day 31, and during this time-frame (i.e., between day 31 and day 42 -- but also before day 31) 1st were able to do anything (though there is no record of what they did). Furthermore, 1st knew that 3rd had been leading with \smamath{608} queries since day 12: with such knowledge, why would 1st even consider attempting a ``model replication attack'' on day 41? Finally, the second submission of 1st (on day 42) occurred only 7 hours after the last submission by 2nd (on day 41), and this submission included 9 webpages that all achieved perfect evasion. Hence, we find it hard to believe that 1st ``started working on the competition'' only after noticing that the leader had \smamath{343} calls (but of course we cannot be 100\% certain).

Secondly, we observe that 2nd arrived at their perfect evasion after knowing that they needed fewer than \smamath{608} queries (the leading score, from 3rd). Hence we conjecture that without the efforts of 3rd, it is unlikely that 2nd would have reached their perfect evasion in \smamath{343} queries; 2nd knew they were constrained to at most \smamath{607} queries and hence had to be more careful. 
Next, 2nd made their first submission on day 31 and their last on day 41, taking at least 19 days to reach the same result that 3rd had achieved in 12 days (although 2nd required roughly half the queries). We can hence conclude that ``3rd required less time than 2nd to accomplish the attack.'' 

Finally, let us go back to 1st; specifically, we focus (again) on their public statement~\cite{phishing1st}. They stated that they wanted to do a model replication (which we can conjecture required ``less effort''), but they changed their mind after seeing the score of 2nd, who themselves achieved after the efforts of 3rd.

\textbf{Conclusions.}
By connecting all these observations, we conclude that without the effort of 3rd, the perfect evasions by 1st and 2nd (with \smamath{320} queries and \smamath{343} queries, respectively) would (likely) not have been achieved. Alternatively, had the leaderboard not been publicly visible, 1st may have opted for a model replication attack (as they themselves stated!). We do not know what 2nd would have done, but 3rd would (likely) still have used their domain expertise to reach their solution---in only 12 days, and requiring \smamath{608} queries (3rd was the first contestant to reach a perfect evasion).

From a different perspective: model replication attacks require more queries (as clearly shown by 4th place), but may require less human effort (as can be extrapolated from 1st actions and statement).

\subsection{Case Study: Malware Competition (2020)}
\label{sapp:cs_malware}
We extend our set of case studies with an in-depth analysis of the Malware Competition organized in the 2020 edition of MLSEC. Here, we elucidate how ``competition-grade'' ML malware detectors can be both hardened and evaded.

\textbf{Rules.} 
This competition had two phases, each with a specific challenge: the first focused on defense, the second on attack. 
Let us describe what each challenge entailed.
\begin{itemize}
    \item (Defense) In the first phase, participants were asked to submit ML systems for malware detection. Submitted solutions had to meet some performance metrics: at most 1\% false positive rate, at least 90\% true positive rate, and at most 5 seconds response time---all of which were measured on an \textit{unknown} test set. Winners were determined depending on how well their solutions performed against the attacks launched in the second phase.
    
    \item (Attack) In the second phase, participants (which could be different from those of the first phase) were given 50 portable executable (PE) malware samples and allowed to manipulate them in any way that preserved the original malicious functionality. The manipulated PE had to evade as many ML-based detectors as possible---including those accepted in the previous defense phase. These ``attackers'' were given API access to the detectors, which resembled \mls{closed} ML systems: attackers could only submit an input and observe the output (i.e., the \textit{probability} that the sample was malicious). Winners were determined on the basis of how each of the 50 manipulated malware samples performed against \textit{all} the considered detectors.
\end{itemize}

\textbf{What did the winners do?}
Let us focus our attention on the methodologies adopted by the winners of each challenge. 
\begin{itemize}
    \item (Defense) Quiring et al.~\cite{quiring2020against} won the defensive challenge. Their solution integrated typical approaches to counter malware, such as signature-matching and ML-based detection based on static analysis; surprisingly, however, they also integrated a defensive layer for \textit{anticipating} adaptive attackers querying the detector---i.e., the actions of the participants in the second challenge. The solution of~\cite{quiring2020against} stopped 77\% of the malware submitted during the attacker's challenge.
    
    \item (Attack) Ceschin et al.~\cite{ceschin2020no} won the attacker challenge by inducing all of their samples to evade \textit{all} the considered detectors, including the winning submission by Quiring et al.~\cite{quiring2020against}. Notably, the strategy adopted by~\cite{ceschin2020no} was \textit{simple}: they observed that ``embedding the malware payload into another binary eliminates most detection capabilities presented by the models.'' Indeed, to quote from~\cite{ceschin2020no}: ``[although] there are approaches for adversarial attacks generation based on complex techniques [...] we show that it is possible to generate attacks using known, simple techniques.''
\end{itemize}

\begin{cooltextbox}
\element{Observation:} Bypassing some ML systems may not require sophisticated techniques (e.g., those based on gradients) commonly adopted in research papers.
\end{cooltextbox}

\subsection{Example: Gains and Losses is not a zero-sum game}
\label{sapp:game}

The gain of an attacker may not correspond to the loss of a defender. Consider, for example, a company \smacal{D} whose main business is the development of ML-based malware detectors and who sells their products to a customer \smacal{C}. Now consider an attacker \smacal{A} that successfully evades one of the detectors developed by \smacal{D} and manages to inject some ransomware to the systems owned by customer \smacal{C}, asking for a ransom in the amount of \smabb{R}. 
Two scenarios can occur:

\begin{itemize}
    \item \textbf{Pay.} If \smacal{C} pays the ransom, then \smacal{C} loses \smabb{R} while \smacal{A} gains \smabb{R}. However, \smacal{D} (the developer of the `evaded' detector) loses very little. In most cases, the contract between \smacal{D} and \smacal{C} protects \smacal{D} in case of ``faults'' in their systems. In contrast, if the contract determines that \smacal{D} is liable for the losses incurred by \smacal{C}, then \smacal{D} will pay \smabb{R} to \smacal{C}; however, in this case, \smacal{D} is typically insured by other companies, who will reimburse \smabb{R} to \smacal{D}. Hence, if the ransom is paid to \smacal{A}, then \smacal{A} gains \smabb{R}, and \smacal{D} loses nothing (assuming that \smacal{C} regains control of their systems).
    
    \item \textbf{Recover.} If \smacal{C} does not pay the ransom, then \smacal{A} gains nothing (actually, \smacal{A} may lose something, e.g., the time spent to carry out the attack). However, \smacal{C} will incur a loss, because their systems are compromised. Two cases can follow, depending on the contract terms:
    \begin{itemize}
        \item If \smacal{D} is liable, then they must help \smacal{C} to recover their data, and \smacal{D} themselves may either incur a loss (e.g., human labour), or not (e.g., if they are insured).
        \item If \smacal{D} is not liable, then \smacal{D} may \textit{gain} something, for example if \smacal{C} asks (and pays) \smacal{D} to assist in the recovery of their systems.
    \end{itemize}
\end{itemize}
Simply put, when looking at cybersecurity from the economic (and operational) viewpoint there are many nuances that can influence the decision to deploy a countermeasure in an ML system. Such considerations further complicate the decision of whether to develop a countermeasure in the first place. The latter decision is made by the system designer, whose main goal is to benefit their bottom line and for whom customer considerations are secondary. (Of course, instilling trust and attracting future customers by having the best system is always an important consideration!)

\subsection{Unreproducible Papers}
\label{sapp:unreproducible}

\textbf{Unavoidable truth.}
The ideal of a publication that releases all its source code and uses public datasets containing attacks from the real world is difficult to meet in practice,  especially in cybersecurity.
The most common and understandable constraint is that of user privacy: real-world data is tied to consumers or employees, and custodians of this data must respect its confidentiality. The employers of security and ML practitioners may also impose constraints with regard to releasing source code, e.g., to avoid giving attackers complete vision of an ML system or a portion thereof. When constraints of this nature do not exist, researchers and practitioners must provide both descriptive details and technical artifacts that maximize reproducibility.

\textbf{What can be done.}
Publications that face privacy constraints and/or cannot release their source code can, however, still be valuable so long as they work within those constraints to make reproducibility a priority. To this purpose, we propose the following \textit{actionable} principles, which can be adopted by research papers to improve their scientific value:
\begin{enumerate}
    \item ML systems and models should be \textit{described with sufficient clarity} that a reader can re-create the model with a high degree of accuracy after reading the paper. Features used by the model should be described in detail.  
    \item Practitioners should \textit{evaluate their models on public datasets} whenever data that is sufficiently similar to the private data is available. This recommendation need not prevent additional evaluation on private datasets.
    \item Papers should \textit{describe as many of the nuances of private datasets as possible} (e.g., examples of feature values and their distributions, correlation between features, and other properties of the data) while not providing individual data points when doing so would violate user privacy. 
    \item \textit{Attacks} that appear in private datasets should be \emph{released when possible}. If they are not releasable, they should be described with sufficient detail to enable researchers to reproduce similar attacks in their own studies and to understand attacker threats that appear in real systems.
\end{enumerate}
We observe, however, that providing all such details can be tough in a research paper: as highlighted in~\cite{apruzzese2022role}, some venues (especially conferences) have a strict page limit. Hence, \textit{we endorse conference organizers to remove such limits} when they are used to provide additional technical details that are not part of the paper's contribution but which are crucial to provide at least some form of reproducibility.

\subsection{Discussion (with the Reviewers)}
\label{sapp:discussion}
\noindent
For informational purposes, we record here some of the discussion we had with our paper's (anonymous) reviewers, since other readers may have similar questions. Transcripts have been lightly edited for clarity.

\subsubsection{Spammers and regular users}
\label{ssapp:spammers}
According to the description in §\ref{ssec:cs_david}, the malicious user wishing to upload a problematic image seems to perform a very similar action to millions of users uploading any sort of images every day on any OSN. Hence, although some malicious user’s action pattern may appear different from those of normal users (especially when automated tooling is used to perform such actions) this is far from being an universal truth. 

\textit{Our Response:}
This statement is correct, and we will happily clarify. Ultimately, ``professional spammers'' are trying to make money. They may do this by, e.g., posting links to counterfeit goods; ``phishing'' legitimate users and selling their account credentials; engaging in 419 scams~\cite{edelson2003419}; or in other ways. Since the expected value of a single spam post is very low (most people don’t buy fake Ray-Bans!), spammers need to post \textbf{a lot} to reach their goal. So they necessarily need to turn to techniques---usually, but not always, automation---that allow them to distribute content quickly across a wide audience. This type of behavior does not match that of a typical Facebook user. Clearly, low-volume spammers who invest a lot of effort in mimicking the behaviour of legitimate users are harder to detect (with or without ML). However, since (a) such an attack requires a high resource investment; (b) the expected value (to the spammer) of a single spam post is still low; and (c) most (but not all) spammers are economically rational, such attacks will be rare and it's acceptable for the \mls{activity} layer to miss them. (Of course, the spammy content may still be caught by the \mls{application} layer.)

\subsubsection{Defenses vs. limited-knowledge attackers (§\ref{ssec:tm_defense})}
\label{ssapp:defense}
It is true that perfect-knowledge attacks are rare, but we should be very careful to draw a distinction between ``defenses that are robust in a query-only setting'' (a reasonable threat model) and
``defenses that only work because the attacker doesn't know what the defender is doing'' (an \textit{un}reasonable threat model).

\textit{Our Response:} This observation is legitimate. We do not advocate that ``all (defensive) papers must consider attackers without perfect knowledge.'' Clearly, it would be beneficial if all papers proposed defenses that work against perfect-knowledge attackers. Our intention, however, is to emphasize that ``real'' attackers may not have perfect knowledge, and hence we find it positive that \textit{some} papers propose defenses specifically addressed at such attackers. Put differently, we think it is unfair that papers may be rejected because ``the threat model considers an attacker without perfect knowledge'': if the threat is (justified to be) likely to occur in reality, then an effective defense can be beneficial to the real world. We believe there is much to learn from these evaluations as well!

\subsubsection{Domain Expertise}
\label{ssapp:domain}
In §\ref{ssec:cs_hyrum}, the 1st--3rd place teams all reached a perfect evasion by exploiting their domain knowledge. Does assuming attackers with different degrees of domain knowledge describe different threat models?

\textit{Our Response:} Tough question. In this case study, all participants had the same \mls{knowledge}, i.e., they knew \textit{nothing} of the ML system (aside from that it analyzed webpages in the form of HTML). However, clearly, all participants had different ``knowledge'' of the respective domain. Note, however, that such knowledge is impossible to quantify: maybe they all perfectly knew the in-and-outs of HTML, but a participant was simply ``lucky enough'' to make a correct initial guess, which they then leveraged to build the remaining parts of the attack. 
In a sense, this case study (§\ref{ssec:cs_hyrum}) shows that \textit{guessing is a crucial part of a real attack} (and we find promising that increasingly more papers are incorporating guessing in their attacks, e.g.,~\cite{bagdasaryan2021blind, apruzzese2022wild}). 
Yet, we believe that such different degrees of ``domain knowledge'' should not be classified as different threat models. From a security standpoint, it is crucial to envision an attacker that is expert in the general domain. Note, however, that this is different from assuming that the attacker has ``perfect \mls{knowledge} of the target ML system''! Hence, we conclude that the threat model should simply state the \mls{knowledge} of the attacker with respect to the ML system, and then assume that the attacker is an expert in any of the components of such ML system that they are aware of.

\subsubsection{``Adversarial'' in practice}
\label{ssapp:adversarial}
Can you clarify what you meant by ``everything is adversarial'' in §\ref{ssec:adapting}?

\textit{Our Response:} From the perspective of a security practitioner, an event is either ``benign'' or ``malicious''; i.e., either the system is under attack or it is not. Hence, the term ``adversarial attack'' is redundant: if there is an attack, then by definition there is an adversary and the attack is ``adversarial.'' Similarly, a ``threat model'' implicitly assumes that there is ``a threat'' (i.e., an attacker) which is obviously adversarial.

In our conversations, security practitioners are often confused when we refer to ``adversarial attacks.'' From a research perspective this term refers to ``attacks that exploit the vulnerability of ML models to adversarial examples,'' while from a practitioner perspective the focus is just ``attack.'' In other words: the term ``adversarial attack'' implicitly means that some attacks are not ``adversarial'' --- which is illogical. 

To provide examples, sometimes in our exchanges with practitioners (P), the following occurs:
\begin{itemize}
    \item P: ``Our systems are constantly under attack.''
    \item Us: ``Are the attacks adversarial?''
    \item P: ``Well of course they are!''
\end{itemize}
Alternatively:
\begin{itemize}
    \item Us: ``Do you test your systems against adversarial attacks?''
    \item P: ``Are there attacks that are not adversarial?''
\end{itemize}
(An even more awkward situation would arise from asking practitioners, ``Are you more scared of adversarial attacks, or of non-adversarial attacks?'')

We believe that a more precise (i.e., less redundant) usage of the term ``adversarial'' will benefit bridging the gap between research and practice in the context of...adversarial ML!

\subsubsection{Nation-state adversaries}
\label{ssapp:nation}
If we consider nation-state adversaries, is human cost measurable here (or relevant)? 

\textit{Our Response:} It depends. Generally speaking, if a ``nation-state adversary'' is assumed to have unlimited resources, then we agree that it makes little sense to report such a cost measure. However, such (extreme) cases should be reported in the paper when presenting the threat model. For instance, a paper should state ``we assume an attacker that is backed by an entire country, and which is hence willing to invest a large amount of resources to carry out their strategy.'' Explicitly stating this assumption would automatically put such a paper in a different bracket with respect to attacker cost.
As an aside, if such a ``nation-state--sponsored attack'' is found to be ``devastating'' (e.g., it leads the entire ML system to malfunction) then real companies would be more interested in studying the corresponding paper and devising appropriate countermeasures. 
(In a sense, what we have just described is a use case of ``Adapting Threat Models to ML Systems'' presented in §\ref{ssec:adapting}). Nevertheless, in reality, even nation-state adversaries do not have unlimited resources: hence, making such an assumption by default is not very realistic.

\section{State-of-the-Art: literature review}
\label{app:sota}

In this Appendix we describe in more depth our literature review summarized in §\ref{sec:sota}. Given the detailed methodology, the systematic analysis, and the original interpretations, we consider this appendix to be a complementary contribution of our position paper.

\subsection{Methodology}
\label{sapp:sota_method}

Inspired by by~\cite{arp2022dos} and by~\cite{apruzzese2022sok}, we carry out our review by adopting a structured approach split into two phases, \textit{selection} and \textit{inspection}. Both phases involved exchanges of opinions among the authors of this paper, aimed at ensuring fairness in the review process and removing potential sources of bias.

\textbf{1$\mathbf{^{st}}$ phase: Selection.}
We conducted our survey in July and August 2022. During this time frame we performed two steps: we \textit{acquired} the proceedings and \textit{filtered} them.

\begin{enumerate}[label=\roman*)]
    \item \textit{Acquire.} We downloaded the proceedings of the ``Top 4'' security conferences (SEC, NDSS, CCS, SP). We considered the editions of these conferences held in the previous three years, i.e., in 2019, 2020, and 2021. We only consider the papers published in the main event (i.e., we do not consider workshops). Overall, these proceedings contained \smamath{435} papers for 2019, \smamath{470} for 2020, and \smamath{644} for 2021.
    
    \item \textit{Filter.} To identify papers within our scope, we started by considering papers included in those sections of the proceedings that specifically focused on ``machine learning.'' We then extended our search by also performing a summary scan of the titles and abstracts of the remainder of the proceedings, identifying additional candidate papers.\footnote{E.g., the paper~\cite{aghakhani2020malware} is included in the ``Malware 2'' section of NDSS20, but it (also) considers attacks against ML-based malware detectors} At the end of this step, we obtained \smamath{30} papers for 2019, \smamath{35} papers for 2020, and \smamath{66} papers for 2021.
\end{enumerate}

\textbf{2$\mathbf{^{nd}}$ phase: Inspection.}
We analyzed these \smamath{131} papers, with two objectives: \textit{filtering} those papers that, despite being related to ML security, were beyond our scope; and \textit{distilling} the knowledge that we wanted to communicate in our work.

\begin{enumerate}[label=\roman*)]
    \item \textit{Filter.} Our focus is on papers on the security of ML and, in particular, on attacks (and defenses) against ML systems. We have already discussed (see §\ref{ssec:related}) some orthogonal research areas. Upon inspecting our candidate papers, however, we found 3 additional areas to exclude from our main analysis due to assumptions that significantly deviate from our main focus. These three areas are \textit{federated learning}~\cite{li2020federated}, where the corresponding issues (such as byzantine fault tolerance~\cite{cao2020fltrust}) mostly pertain to the distributed systems domain; \textit{robust, efficient, or secure computation}, because they mostly focus on technical implementations and do not propose or consider any adversarial ML attack; and \textit{high-level analyses}, because they do not propose any original attack or defense in the ML security context. Eventually,\footnote{We provide in our website~\cite{radcg:website} the list of 43 papers that we excluded.} we obtained \smamath{23} papers for 2019, \smamath{24} papers for 2020, and \smamath{41} papers for 2021.
    
    \item \textit{Distill.} Finally, we thoroughly analyzed the final set of \smamath{88} papers by going through each paper and answering a set of 12 questions.  Each question focused on deriving knowledge significant for our position paper, as well as for promoting future work in this research field. Although most of our questions can be easily answered, some are not straightforward (as we will discuss). To minimize the likelihood of errors and reduce subjective bias, we carried out our analysis in pairs: two authors independently reviewed each paper and then discussed the findings in a series of meetings. We repeated this process five times. 
\end{enumerate}

Before we delve deeper into our analysis, we provide some interesting trivia.
In 2019, most of the relevant papers were put in sessions entirely devoted to ``ML security.'' In 2021, however, we found sessions entirely dedicated to specific attacks (e.g., ``attacks on speech recognition'' or ``inference'' at SP2021).
Some papers have very similar contributions across the same conference. For instance,~\cite{chen2021real} and~\cite{abdullah2021hear} (both at SP21) propose attacks on speech recognition (and both, surprisingly, evaluate their proposals also on real humans); whereas~\cite{salem2020updates} (SEC20) and~\cite{zanella2020analyzing} (CCS20) propose membership inference attacks that leverage the updates of ML models.

\subsection{Research Questions}
\label{sapp:questions}

We inspected 88 papers by asking ourselves 12 research questions, divided into two groups. 

\textbf{Generic Questions (G).} The first eight (\qg{1} to \qg{8}) are generic, and are meant to provide a broad overview of the latest trends in research---some of which are reported in our main paper (§\ref{sec:sota}). For each of the 88 works, we ask ourselves:
\begin{enumerate}[label={\small G{{\arabic*}})}]
    \item Does the paper \textit{focus} on an attack or on a defense?

    \item What is the main attack \textit{family} (i.e., poisoning, stealing, evasion, membership inference)?
        \item What \textit{paradigm} of the ML model that is subject to the attacks (i.e., does it rely on shallow or deep learning)?
    \item Are the \textit{costs} taken into account (in any way)?
    \item What are the \textit{data-types} (i.e., images, audio, text, or other) considered in the evaluation?
    \item Has the \textit{source-code} been publicly released?
    
        \item Has a complex \textit{pipeline} been reproduced in the evaluation (i.e., does the ML system consist in just an ML model)?
        \item Does the paper consider an ML system \textit{deployed} in the real world (and, if yes, what is its \textit{type})?

\end{enumerate}
The answers to \qg{1} to \qg{8} are shown in Figs.~\ref{figs:g12} to~\ref{fig:real}, all sharing the same structure. Each figure refers to a specific question, and presents four stacked bars, one per year [2019--2021] while the rightmost bar is an aggregate. Each stacked bar reports the answers to the corresponding question, both in terms of relative (y-axis) and absolute (written on the specific bar) frequency over a given stack. Descriptions and interpretations on these results are provided in appendices~\ref{sapp:g12} to~\ref{sapp:g78}.

\textbf{Threat-Model Questions (T).}
The last four (\qt{1} to \qt{4}) relate to the threat model envisioned in the paper. Specifically, we consider the \textit{weakest} (we explain our reason in Appendix~\ref{sapp:table_sota}) attacker assumed in a paper, and then ask ourselves:
\begin{enumerate}[label={\small T{{\arabic*}})}]
    \item Does the attacker know the ML model's \textit{parameters}?
    \item Does the attacker know the \textit{semantics} of the input data fed to the ML model?
    \item Can the attacker observe the \textit{output} of the ML model (and, if yes, \textit{how})?
    \item Does the attacker have any power on the \textit{training set}?
\end{enumerate}
We explain how we answered \qt{1} to \qt{4} in Appendix~\ref{sapp:specific}.

The answers to all our questions for all our considered papers are provided in Table~\ref{tab:sota}, described in Appendix~\ref{sapp:table_sota}.

%%%%%%%%%%%%%%%%%%%%%%%%%%%%%%%
\subsection{Answers to G1 and G2: Focus, and Attack Family}
\label{sapp:g12}

\begin{figure}[!htbp]
    \centering
    \begin{subfigure}[t]{0.45\columnwidth}
        \centering
        \includegraphics[width=1\columnwidth]{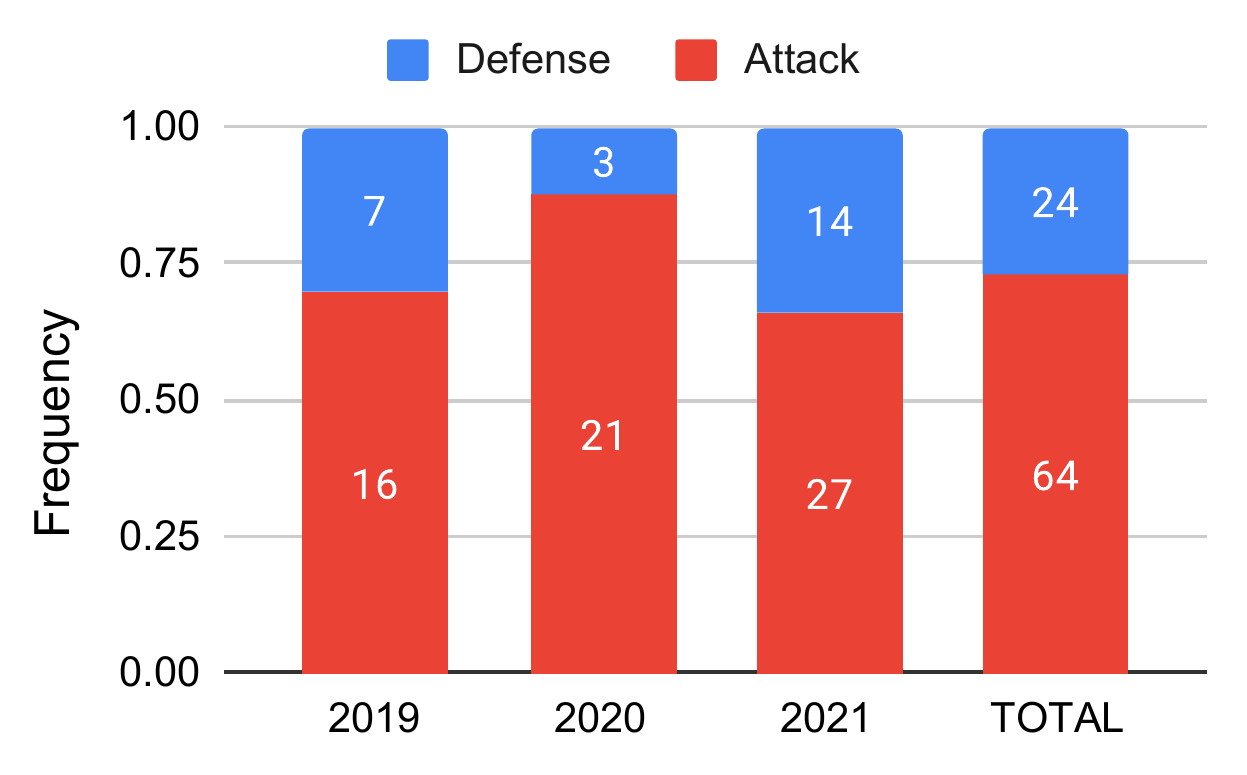}
        \caption{G1: What is the \textit{focus} of the paper?}
         \label{sfig:focus}
    \end{subfigure}%
    ~ 
    \begin{subfigure}[t]{0.45\columnwidth}
        \centering
        \includegraphics[width=1\columnwidth]{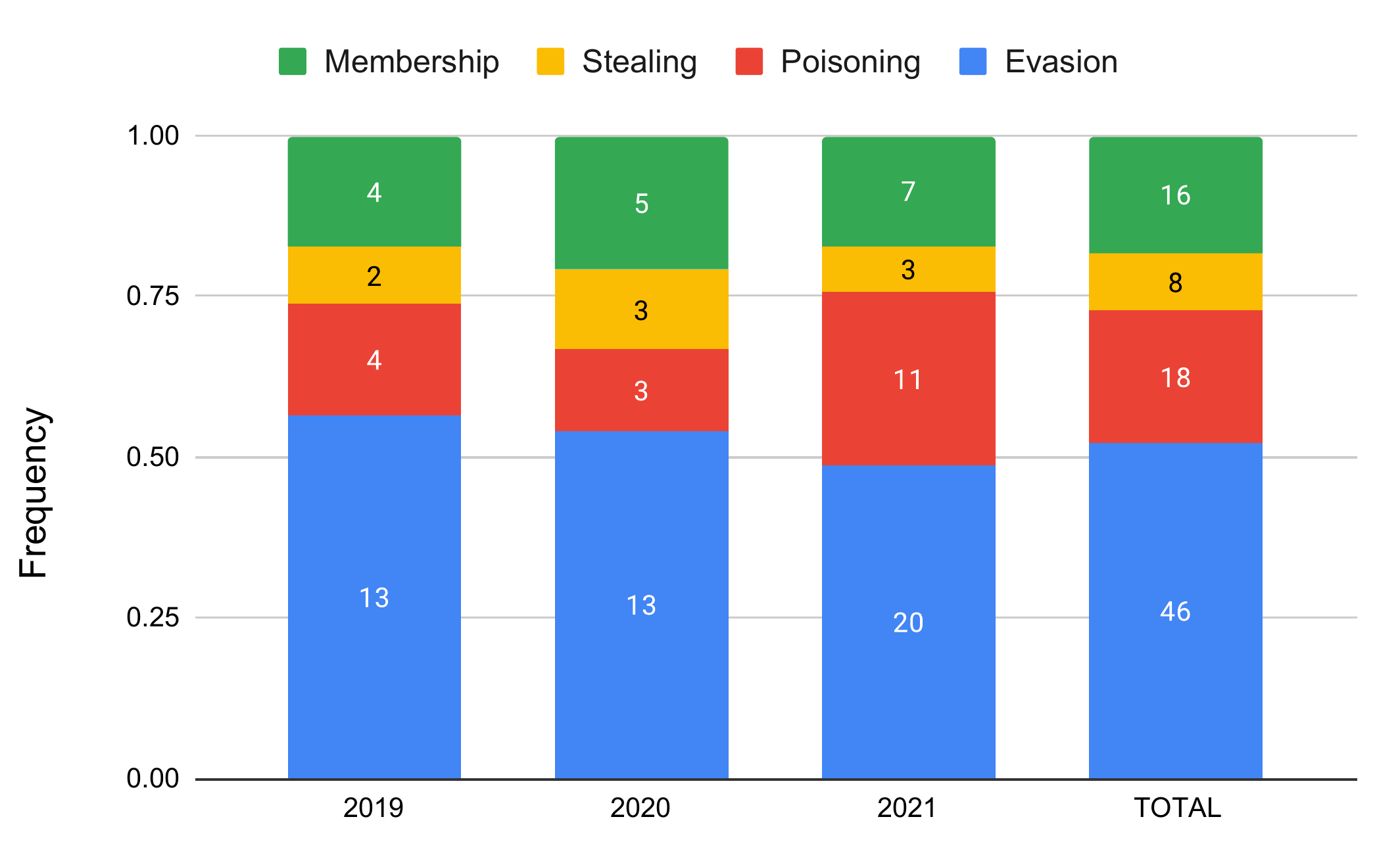}
        \caption{G2: What is the main attack \textit{family}?}
         \label{sfig:family}
    \end{subfigure}
    %\vspace{-3mm}
    \caption{Answers to G1 and G2.}
    \label{figs:g12}
    \vspace{-1em}
\end{figure}

The answers to the first two questions are shown in Figs.~\ref{figs:g12}.

\textbf{\qg{1}: Focus.} 
This question is straightforward to answer: does the paper focus on attacks against ML, or on defenses to attacks against ML? In some rare cases, some papers simply proposed a ML method to solve a given task (e.g., intrusion detection~\cite{jan2020throwing}) but also evaluated this method against some evasion attacks: we considered these as attack papers, because the proposed ML method is not typically designed to withstand adversarial ML attacks.
From Fig.~\ref{sfig:focus}, we can see that 75\% of papers put a stronger emphasis on the attack---although most of these papers (e.g.,~\cite{carlini2021poisoning}) also discuss some countermeasures to the corresponding attack. This result is not surprising, as `attack papers' are typically considered to have a higher novelty value---because the attack stems from a different threat model, or considers a new domain. 

\textbf{\qg{2}: Attack Family.}
Inspired by~\cite{bieringer2022industrial}, we consider four main attack families: \textit{poisoning}, \textit{model stealing}, \textit{evasion}, as well as \textit{membership inference}.
Fig.~\ref{sfig:family} shows that \textit{most papers envision evasion attacks} ($\sim$50\%), whereas poisoning, model stealing, and membership inference attacks are less prominent (20\%, 10\% and 20\%, respectfully). However, we can see a decreasing trend of evasion, and a \textit{rising trend for poisoning attacks}, especially in 2021. We conjecture that this phenomenon is due to the 2020 paper by Kumar et al.~\cite{kumar2020adversarial}, highlighting that industry is particularly worried about poisoning.

\subsection{Answer to G3: ML paradigm}
\label{sapp:g3}

\begin{figure}[!htbp]
    \centering
        \centering
        \includegraphics[width=0.7\columnwidth]{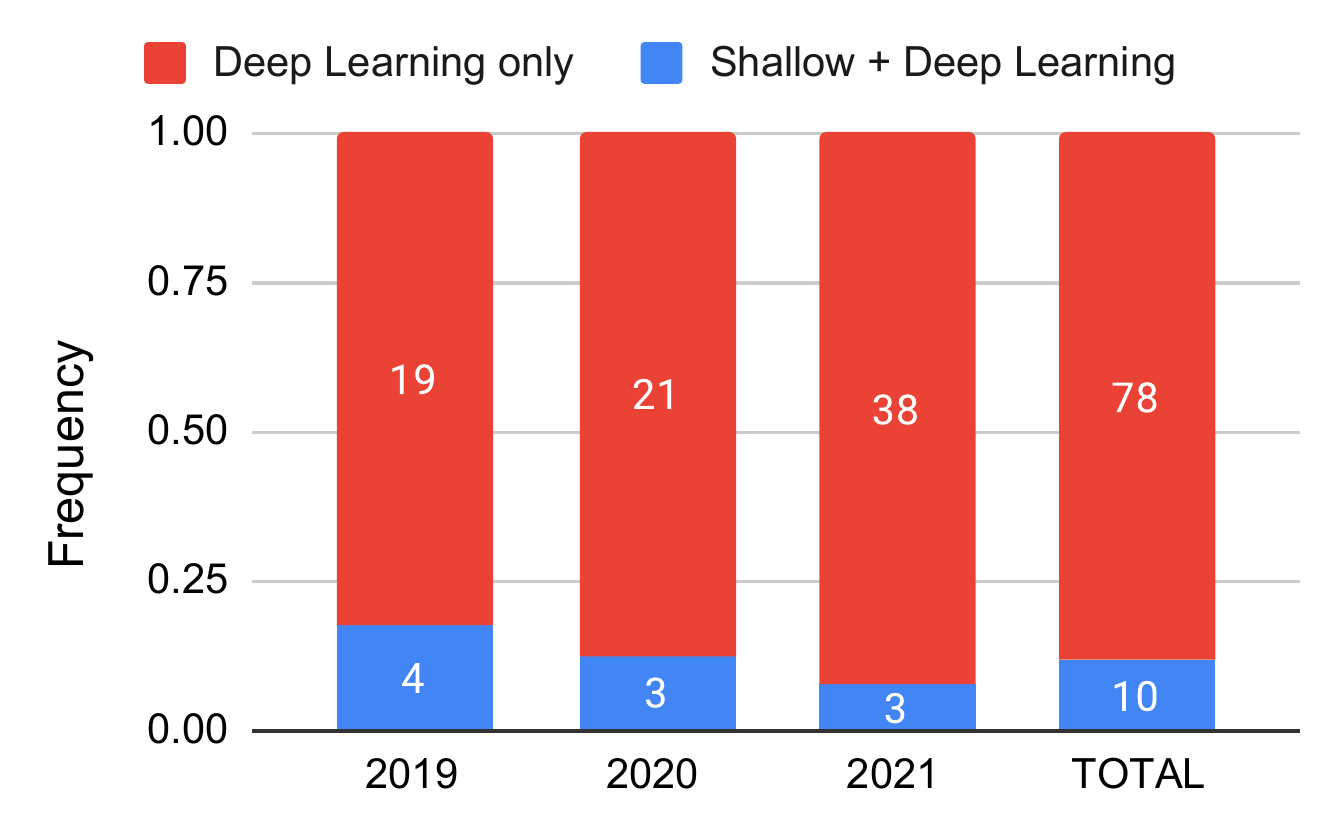}
        \caption{G3: what is the considered ML \textit{paradigm}?}
         \label{fig:paradigm}
    % \vspace{-1em}
\end{figure}

This question is also straightfoward: is the underlying algorithm used by the `targeted' ML model based on shallow or deep learning? The results are in Fig.~\ref{fig:paradigm}.

It is apparent that most papers consider ML systems \textit{exclusively based on DL} (i.e., those using neural networks). Only 10 papers (out of 88) evaluate ML systems based on shallow learning (SL) algorithms (interestingly, two papers consider \textit{only} SL:~\cite{Pierazzi:Intriguing} and~\cite{barradas2021flowlens}). We believe that future efforts should put more focus on SL, for a twofold reason:
\begin{itemize}
    \item \textit{in some domains, SL is better than DL}~\cite{ Apruzzese:Deep, chandrasekaran2020exploring}. E.g., we quote from~\cite{aghakhani2020malware}: ``Even though in our experiments we used SVM, deep neural networks, and different variants of decision-tree learners, like random forest, we only discuss the results of the random forest approach as (1) we observed similar findings for these approaches, with random forest being the best classifier in most experiments, and (2) random forest allows for better interpretation of the results compared to neural networks.''
    
    \item \textit{SL exhibit different properties than DL.} We find insightful to quote a statement by Xu et al.~\cite{xu2021detecting} (emphasis ours): ``we mainly detect AI Trojans on neural networks. We do not include other ML models in our discussion mainly because \textit{there is no current research showing that they suffer from backdoor attacks}.'' Moreover, some SL methods (e.g., Decision Trees) do not employ gradients, meaning that some gradient-based attacks (as well as gradient-based defenses) may not work on them.
\end{itemize}
We also mention that our case study (§\ref{ssec:cs_david}) elucidated that SL methods are used \textit{also} in commercial-grade ML systems.

\subsection{Answer to G4: Cost}
\label{sapp:g4}

\begin{figure}[!htbp]
    \centering
    \includegraphics[width=0.95\columnwidth]{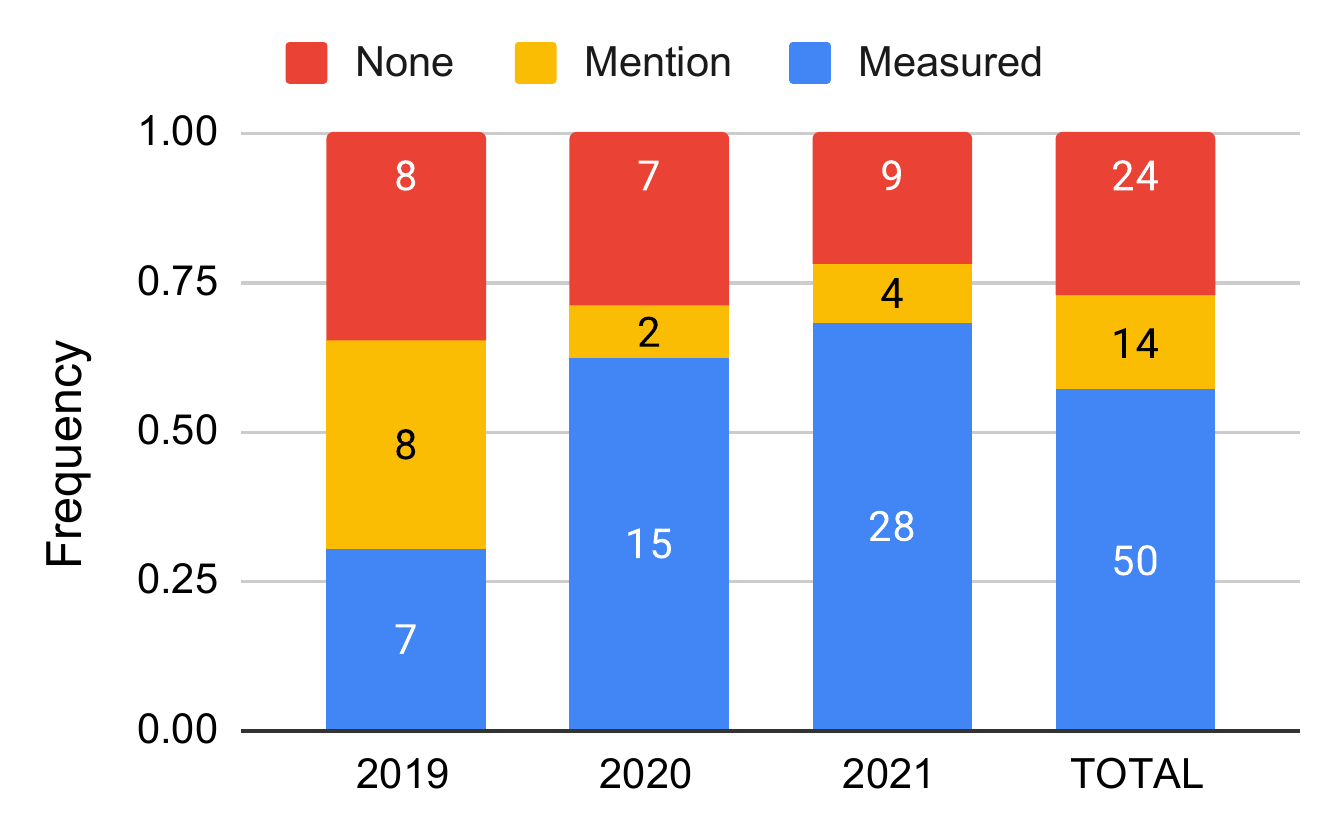}
    \caption{G4: are the \textit{costs} taken into account (in \textit{any} way)?.}
    \label{fig:cost}
\end{figure}

As we discussed (§\ref{ssec:economics}), the \textit{cost} (i.e., the economical factor) plays a crucial role in operational cybersecurity. When analyzing prior work, we considered three possible answers to \qg{4}: whether a paper \textit{measured} the cost in some way (e.g., performance tradeoff, required queries); whether this cost was at least \textit{mentioned} or discussed; and whether \textit{no consideration} on the cost was made. These answers can cover both attack and defense papers. The results are shown in Fig.~\ref{fig:cost}.

These results show a positive trend. In 2019, less than 30\% of the papers did some measurements of the cost, and 30\% made no consideration whatsoever. The prevalence changes in 2020 and 2021, as less than 25\% papers did not take into account the economical factor, whereas more than 60\% performed actual measurements. 

\textbf{Cost: Attack papers.} We highlight three papers which attempted to \textit{quantify} the cost of an attack in real currency.
\begin{itemize}
    \item (Equipment) Lovisotto et al.~\cite{lovisotto2021slap} evade object detectors via artificial light manipulation through \textit{commercial projectors}. Some of these projectors have an MSRP of few hundreds dollars, whereas others are more expensive.
    
    \item (Queries) The authors of~\cite{yu2020cloudleak} and~\cite{chandrasekaran2020exploring} alongside measuring the cost of the proposed attack in terms of API calls (i.e., queries), but also quantified the actual cost required to make these API calls. Specifically, they considered the \textit{prices of popular MLaaS}, and derived that their attacks only necessitate few dollars.
\end{itemize}

\textbf{Cost: Defensive papers}. The most typical way to measure the cost in defensive papers is via the `tradeoff', i.e., the change in performance \textit{after} applying a given countermeasure (e.g.,~\cite{li2020textshield}). We mention two works that make insightful considerations on the `defensive' cost---or rather, on how the principle of a defense is to increase the `cost' of the attacker.
\begin{itemize}
    \item Chen et al.~\cite{chen2021cost} propose to measure the cost of feature manipulation against shallow ML models. The intuition is to carry out the measurements during the \textit{evaluation} of ML models, with the underlying principle of ``if an attacker wants to evade an ML model by doing $a$, they have to spend $c_a$; and if they want to do $b$, they have to spend $c_b$.'' We remark, however, that~\cite{chen2021cost} does \textit{not} evaluate any attack: the main contribution is measuring the (computational) effort in manipulating some features. (which is why we do \textit{not} include~\cite{chen2021cost} in our Table~\ref{tab:sota}.)
    
    \item Heisenhofer et al.~\cite{eisenhofer2021dompteur} acknowledge that attackers can use adversarial examples to evade a ML system for automated speech recognition. Hence, they aim to increase the cost of the attacker to craft such adversarial examples.
\end{itemize}
An alternative formulation of the ``raise the attacker's cost'' is intrinsic of papers that focus on the \textit{detection} of security violations. We observe, however, that this detection can occur at different \textit{granularities}. For instance, some papers aim to detect individual adversarial examples (e.g.,~\cite{hussain2021waveguard}) or poisoned ML models (e.g.,~\cite{tang2021demon}); whereas others have a broader scope and aim to detect attackers (e.g.,~\cite{shan2020gotta}).

\subsection{Answers to G5 and G6: Data-type and Source-code}
\label{sapp:g56}
We now turn the attention to \qg{5} and \qg{6}, which are strongly related to the \textit{reproducibility}~\cite{hutson2018artificial} of research. We make a statement on this subject at the end of our main paper (§\ref{ssec:disclosure}).

\begin{figure}[!htbp]
    \centering
        \includegraphics[width=0.85\columnwidth]{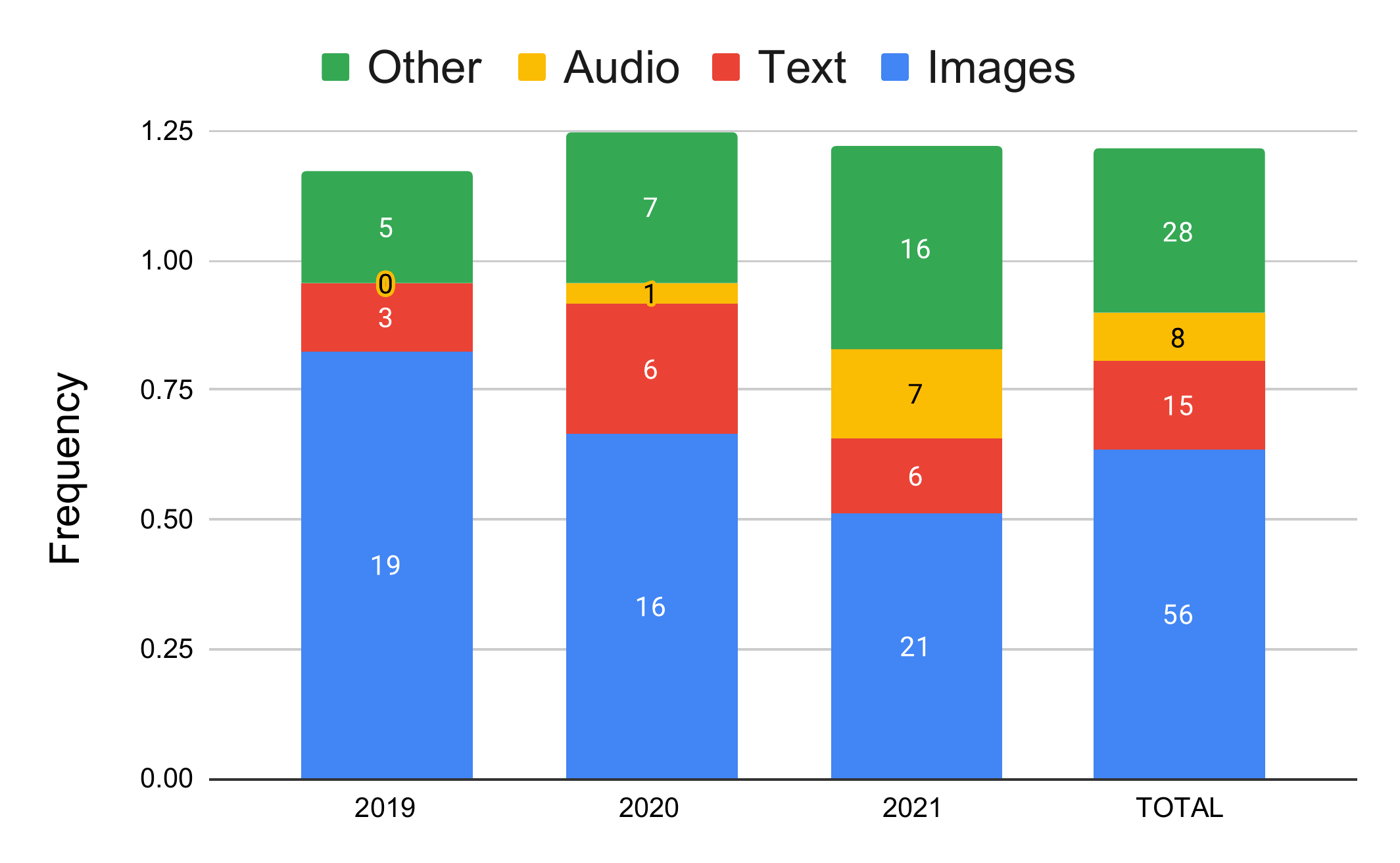}
        \caption{What are the \textit{data-types} considered in the evaluation?}
    \label{fig:data}
    % \vspace{-1em}
\end{figure}

\textbf{G5: Data-type.}
Past research investigated a plethora of application domains of ML, which we divide in four categories. Three correspond to the broad areas of \textit{images} (e.g., computer vision), \textit{text} (e.g., natural language processing), \textit{audio} (e.g., automated speech recognition); the fourth includes all \textit{other} domains (e.g., malware). These results are in Fig.~\ref{fig:data}; some papers (e.g.,~\cite{song2021systematic}) consider multiple domains, which is why the y-axis goes above 1. More details are shown in Table~\ref{tab:sota}.

We can see an interesting trend from Fig.~\ref{fig:data}: in 2019, over 75\% of the papers focused on images (typically using well-known datasets such as MNIST and CIFAR), whereas the percentage dropped to 50\% in 2021. The audio domain has also been neglected in 2019, but became prominent in 2021. In particular, papers of 2021 covered much more `novel' domains, such as graphs~\cite{xi2021graph} or games~\cite{wu2021adversarial}.

\begin{figure}[!htbp]
    \centering
        \includegraphics[width=0.85\columnwidth]{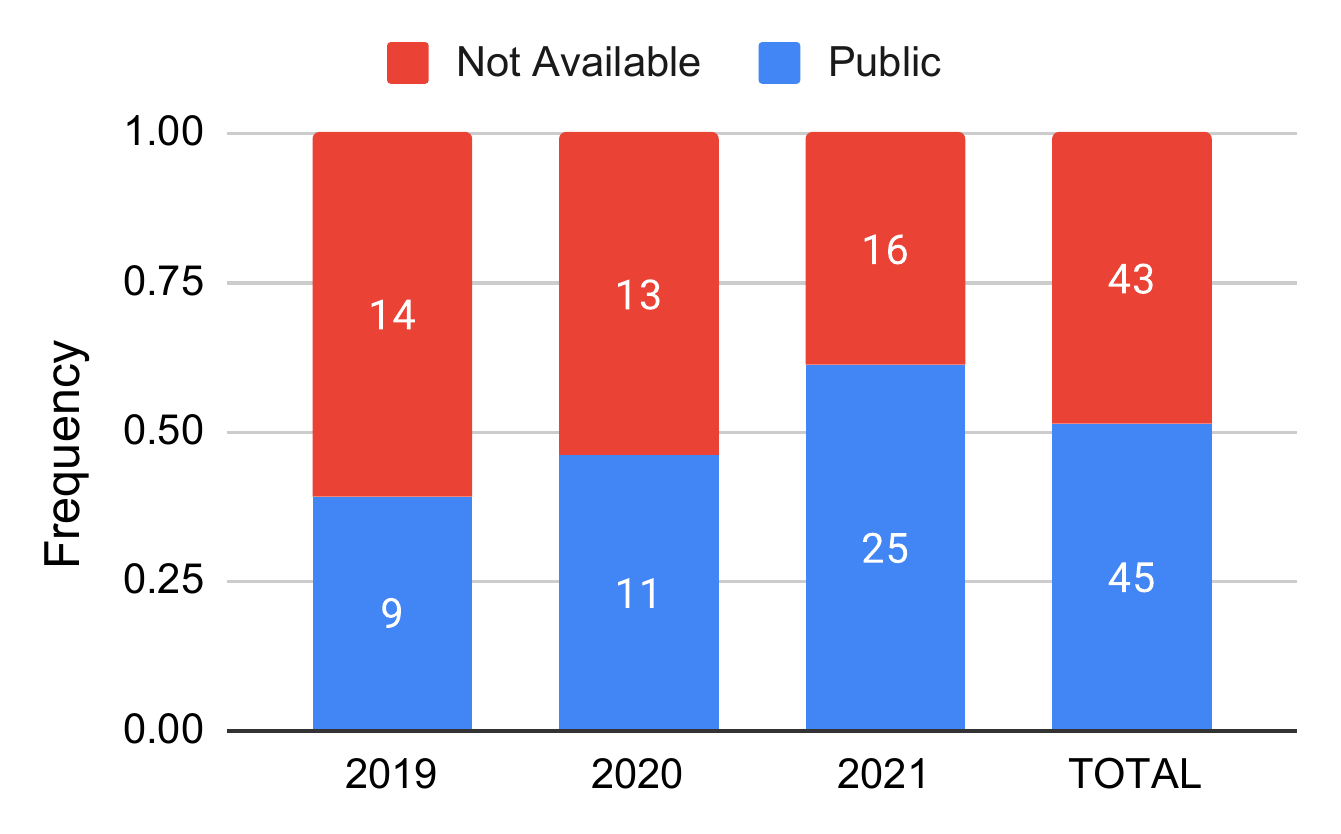}
        \caption{Has the \textit{source-code} been publicly released?}
    \label{fig:code}
    \vspace{-1em}
\end{figure}

\textbf{G6: Source-code.} We looked for links pointing to the source-code of the experiments carried out in all our considered papers. The results are in Fig.~\ref{fig:code}. 

We can see an encouraging trend, as 75\% of papers did not release their source-code in 2019, whereas only 35\% did not do so in 2021. To facilitate future research, we provide also the actual links to every open code repository of each paper in Table~\ref{tab:sota}. Nonetheless, we stress that even those papers that do not make their evaluation publicly accessible had \textit{valid reasons} for doing so. For instance,~\cite{carlini2021poisoning} describes plenty of details, uses well-known public datasets, and mostly rely on \textit{existing} and publicly accessible code. On the other hand, the experiments in~\cite{jan2020throwing} are based on \textit{sensitive data} that cannot be disclosed.

\subsection{Answers to G7 and G8: pipeline and type of ML system} 
\label{sapp:g78}

The last generic questions pinpoint the degree of `realism' considered in the evaluations of prior work. Answering these two questions was not simple: the authors who took part in this activity had several debates before reaching a consensus.

\begin{figure}[!htbp]
        \centering
        \includegraphics[width=0.85\columnwidth]{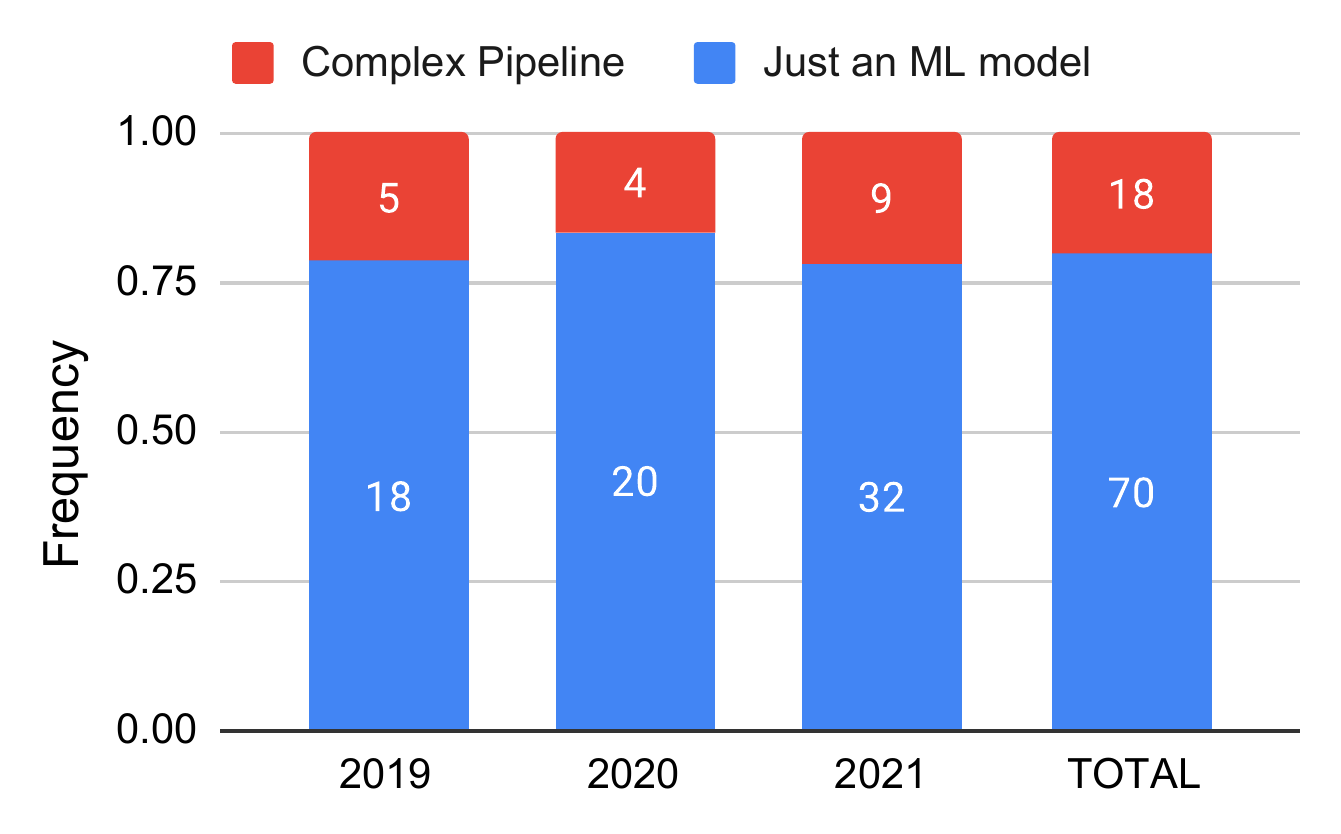}
        \caption{Has a complex \textit{pipeline} been reproduced in the evaluation?}
         \label{fig:pipeline}
         %\vspace{-1em}
\end{figure}

\textbf{G7: processing pipeline.}
This question was among the hardest to answer during our survey. Our goal was determining whether a given paper (i)~envisioned a ML system, and (ii)~developed more components of the ML system than just an ML model. For example, if a paper considered a ML system (by, e.g., providing a schematic), but the evaluation spanned over a single ML model, the answer to \qg{7} was negative. The answer was also negative if the paper made some preliminary checks to the inputs of the ML model (e.g., to ensure that the samples do not violate domain constraints~\cite{sheatsley2021robustness}). Simply put, to answer \qg{7} we embraced Abdullah et al.~\cite{abdullah2021hear} statement, which we quote: ``Processing pipelines for modern systems are comprised of signal preprocessing and feature extraction steps, whose output is fed to a machine-learned model. Prior work has focused on the models, using white-box knowledge to tailor model-specific attacks.'' Answers to \qg{7} are in Fig.~\ref{fig:pipeline}.

\vspace{-2mm}

As expected, nearly 80\% of papers perform their attacks by considering a single ML model. Even some papers that attack MLaaS (e.g.,~\cite{salem2019ml}) ultimately simply have the MLaaS train a given ML model, and then attempt to violate this model without reproducing a real pipeline. Nonetheless, there are some notable efforts, such as~\cite{Pierazzi:Intriguing} and~\cite{xiao2019seeing}. These works reproduce a custom ML system by developing \textit{also} a preprocessing component: this component represents an additional layer that an attacker must bypass to reach the actual ML model that will be ultimately targeted. However, we find it concerning that the overall trend shown in Fig.~\ref{fig:pipeline} is \textit{stable}. We strongly endorse future work to start developing more components

\begin{figure}[!htbp]
        \centering
        \includegraphics[width=0.85\columnwidth]{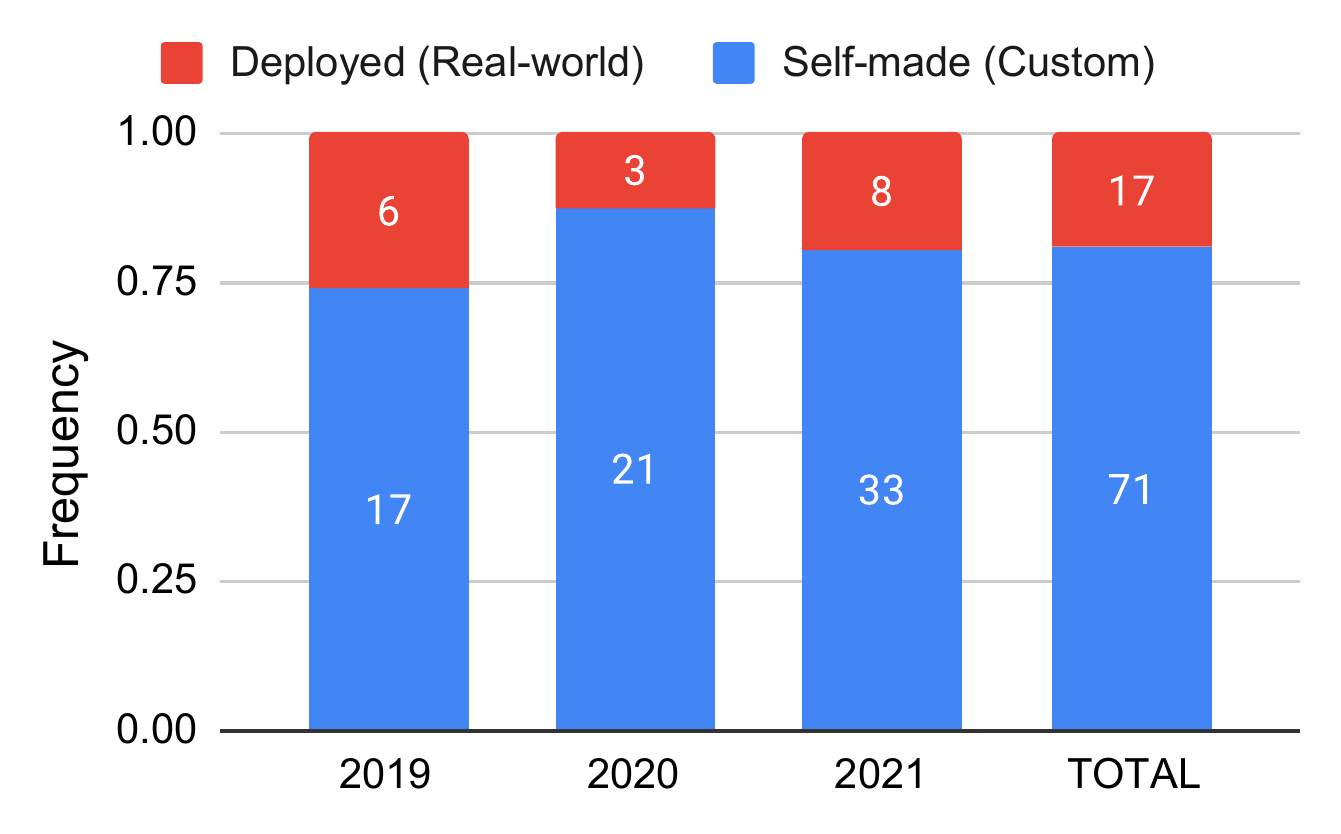}
        \caption{Does the paper consider an ML model \textit{deployed} in the real world?}
         \label{fig:real}
    \end{figure}

\textbf{G8: type of ML system.}
This question is split in two parts. First, we asked ourselves whether, in any part of the evaluation of a given paper, there was an ML model that can be considered as `deployed in practice'. For example, papers that attack commercial ML systems (e.g.,~\cite{nassi2020phantom}) automatically fall into this category; however, we also considered papers that used `extremely popular' ML models (e.g., Yolo~\cite{yolo}) that have been shown to have realistic applications besides benchmarking. The answers to the `first' part of \qg{8} are shown in Fig.~\ref{fig:real}. Conversely, if a paper used some `popular' techniques as a form of preprocessing but the attacked ML model is trained on benchmarks, then the answer to our question was negative---this is the case, e.g., of~\cite{Pierazzi:Intriguing}.
Then, if the answer to the first part of \qg{8} was positive (i.e., the paper attacked a `deployed' ML system), we proceeded to identify the type of (real) ML system. Recall (§\ref{ssec:mls}) that there exist two main types of ML systems: \mls{open} and \mls{closed} (and also \mls{invisible}). The answer to the second part of \qg{8} is in Table~\ref{tab:sota}.

\vspace{-2mm}

By observing Fig.~\ref{fig:real}, we are not surprised to find out that over 80\% of the papers attack ML models (or ML systems) that are entirely self-developed---either by using private data, or by using well-known benchmark datasets. We conjecture that this is due to the difficulty of researchers to acquire permission to use `commercial' ML systems for research. For example, the notable work by Nassi et al.~\cite{nassi2020phantom} clearly explains what the authors had to do in order to attack the ML system for object recognition integrated in real autonomous cars. We commend the work by Nassi et al.~\cite{nassi2020phantom}, but similar experiments may be ``outside the reach'' of most researchers.

Nevertheless, a detailed look at Table~\ref{tab:sota} reveals that, among those papers that attack real ML systems, there is not a single one which is \mls{invisible}. This makes sense: such systems are typically protected by NDA and researchers can hardly use them for experimental purposes. We hope that our position (§\ref{ssec:collaboration}) will promote practitioners to collaborate more actively with researchers, so that future works can assess the robustness of real ML systems against against security violations.

\subsection{Answers to the Threat-Model Questions}
\label{sapp:specific}

There is a remarkable lack of clarity in the way that research papers present their threat models (as also hinted in §\ref{ssec:tm_confusing}). 

\textbf{T1: Parameters.} 
We focused on determining whether the attacker knew anything about the ML model itself, such as the underlying algorithm (e.g., a Decision Tree or a Neural Network), its initial configuration (e.g., the learning rate or the architecture of the neural network) or its learned weights. However -- for those papers in which the attacker was assumed to have some knowledge on the ML model -- it was difficult to specifically determine what was known. Hence, we considered three possible answers to this question: \textit{full}, \textit{zero} and \textit{intermediate} knowledge. An exemplary case of the latter is the work by Tang et al.~\cite{tang2021demon}, stating that their (``black-box'') attacker: ``does not have information about the inner parameters of the target model [but] he knows the target model’s architecture, used optimization algorithm and hyper-parameters.''.

\textbf{T2: Semantics.} 
We consider whether the attacker was aware of the input data-type received by the ML model. For instance: does the attacker \textit{know} that the ML model analyzes images/malware/text? The answer was binary: \textit{yes} or \textit{no}. In most cases, the answer was positive. However, in some rare occurrences the attacker was also oblivious of this information. An example is~\cite{hong2019terminal}, in which the attacker simply does not need this knowledge because the manipulation affects the ML system at the hardware layer.

\textbf{T3: Output.} 
The main difficulty we encountered when answering this question was the \textit{format} in which the attacker received the output of the ML system. As highlighted by Jagielski et al.~\cite{jagielski2020high}, an attacker can receive many forms of feedback by a given ML system. For simplicity, we considered four answers: \textit{decision} (e.g.,~\cite{tramer2019adversarial}) if the attacker can only observe the decision of the whole ML system; \textit{label} (e.g.,~\cite{li2021membership}), if the attacker received the class with the highest probability; \textit{probability}, if the attacker received any additional information beyond the label (e.g.,~\cite{zhao2019seeing}); as well as \textit{none}, when the attacker did not receive any feedback (or was not needed).

\textbf{T4: Training.} 
As stated in our main paper (see §\ref{ssec:tm_confusing}), identifying if the attacker had some knowledge on the training data was very confusing. To be as specific as possible, we considered the following answers: \textit{full}, if the attacker had full access (read and write) to the training data; \textit{read}, if the attacker can observe the entire training set; \textit{subset} (e.g.,~\cite{yao2019latent}, if the attacker had a subset of the actual training dataset; \textit{surrogate} (e.g.,~\cite{salem2020updates}), if the attacker had samples not included in the training data, but having the same distribution; \textit{distribution} (e.g.,~\cite{li2018textbugger}), if the attacker only knows the class distribution of the training data; and \textit{none} if the attacker knows nothing.

\textbox{Remark}{Quantitative analyses that take into account the cumulative answers to \qt{1} to \qt{4} are not possible: every paper ultimately considers a different scenario. Therefore, we refrain from deriving any `trend' from \qt{1} to \qt{4}. Considerations should be made on a paper-by-paper basis, and we discuss some of them in the main paper (§\ref{ssec:tm_attack} and §\ref{ssec:tm_defense}). The individual answers to these questions are provided in Table~\ref{tab:sota}.}

\subsection{Complete Table}
\label{sapp:table_sota}
We summarize our complete literature review in Table~\ref{tab:sota}. This table\footnote{We note that some (actually, most) papers envisioned attackers conforming to diverse threat models. Due to some of statements made in our main paper (\ref{ssec:cost_tm}), we show in Table~\ref{tab:sota} the assumptions that correspond to \textit{weakest} attackers (e.g., if a paper considers both a black- and white-box attacker, we will report the description of the black-box attacker in Table~\ref{tab:sota}). The motivation is that attacks stemming from weaker adversaries tend to be more representative of the real world. Moreover, some papers consider dual attacks (e.g.,~\cite{Demontis:Adversarial} and~\cite{pang2020tale}). In these cases, we only report the one that is given a greater emphasis in the paper.} lists each of the 88 papers considered in our analysis, presented in temporal order: from 2019 until 2021, and from NDSS (typically held in January) until CCS (typically held in November). For each paper, we report the first author name, as well as concise answers to our 12 questions.
    
    \textbf{Generic Questions (\qg{1} to \qg{8}).} 
    Most of these are straightforward. For readability, every `negative' answer is denoted with a blank cell. For \qg{1} (focus), an asterisk denotes a paper whose main contribution is a ML-based solution to a given problem, which is \textit{also} evaluated in adversarial scenarios (e.g.,~\cite{aghakhani2020malware}). For \qg{4} (cost), we use: \xmark{} when no consideration is made; \halfcirc{} when it is just mentioned; and \cmark{} when some measurement is performed. For \qg{5}, we also provide the specific data-type (e.g., malware) in the `other' column (if there are more than one `other' data-type, we report `+'). For \qg{6} (source code), the \cmark{} is a \textit{link} that leads to the actual repository.  For \qg{7}, we use \cmark{} if the paper develops an actual pipeline, i.e., at least another component besides the ML model (an asterisk denotes attacks at a different layer of the pipeline---e.g., hardware~\cite{hong2019terminal}).
    
    \textbf{Threat-Model Questions (\qt{1} to \qt{4}).} 
    For \qt{1} (parameters), we use \xmark{} if nothing is known, \cmark{} if the ML model is fully known, and \halfcirc{} if the attacker knows a portion of its architecture, or has a surrogate model. 
    For \qt{11} (output), we use $l$ if it is the actual label, $p$ if it is a probability, $s$ if it is the decision of the whole ML system, and \cmark{} if it is not needed. 
    For \qt{11} (training data), we use: \cmark{} if the attacker has complete read and write access; $R$ if they have complete knowledge of the training data; $\subset$ if the attacker has a subset of the training data; and $S$ if they have a surrogate dataset of the same distribution; and $D$ if they only know the distribution of the training set; \xmark{} if they know (and can do) nothing about the training data. However, if the answer to \qt{2} is `poisoning', it is implicitly assumed that the attacker always has some form of write access to the training data (the only exception is~\cite{bagdasaryan2021blind}).

\begin{table*}[htbp]
    \centering
    \caption{The 88 papers considered in our analysis. Each column reports the answer to one of the 12 research questions we used during our survey. If available, the \textbf{G6} column provides the \textit{hyperlink} to the websites hosting the source-code of a given paper. Explanations are in Appendix~\ref{sapp:table_sota}.} 
    \label{tab:sota}
    % \vspace{-3pt}
    \resizebox{1.95\columnwidth}{!}{
        \begin{tabular}{c|c|c?c|c|c|c|c|c|c|c|c|c|c?c|c|c|c}
            \toprule
            
            \multirow{2}{*}{\begin{tabular}{c} \textbf{Year} \\ (subs)  \end{tabular}} &
            \multirow{2}{*}{\begin{tabular}{c} \textbf{Venue} \\ (subs)  \end{tabular}} &
            \multirow{2}{*}{\begin{tabular}{c} \textbf{Paper} \\ (1st author)  \end{tabular}} & 
            \multicolumn{1}{c|}{\textbf{G1}} & 
            \multicolumn{1}{c|}{\textbf{G2}} &
            \multicolumn{1}{c|}{\textbf{G3}} & 
            \multicolumn{1}{c|}{\textbf{G4}} &
            \multicolumn{4}{c|}{\textbf{G5} (Evaluation Data)} &
            \multicolumn{1}{c|}{\textbf{G6}} & 
            \multicolumn{1}{c|}{\textbf{G7}} & 
            \multicolumn{1}{c?}{\textbf{G8}} &
            \multicolumn{1}{c|}{\textbf{T1}} &
            \multicolumn{1}{c|}{\textbf{T2}} &
            \multicolumn{1}{c|}{\textbf{T3}} &
            \multicolumn{1}{c}{\textbf{T4}} \\ \cline{4-18}

            & & & \textit{Focus} & \textit{Attack} & \textit{Paradigm} & \textit{Cost} & \textit{Img} & \textit{Text} & \textit{Audio} & \textit{Other} & \textit{Code} & \textit{Pipeline} & \textit{Type} & \textit{Param.} & \textit{Sem.} & \textit{Output} & \textit{Training} \\

            \midrule
            
            \multirow{23}{*}{\begin{tabular}{c} 2019\\ {\tiny (23/435)}            \end{tabular}} & 
            \multirow{4}{*}{\begin{tabular}{c} NDSS \\ {\tiny (4/89)}            \end{tabular} }
            & Salem~\cite{salem2019ml} & atk & {\scriptsize Member.} & DL & \halfcirc & \cmark & \cmark &  &  & \href{https://github.com/AhmedSalem2/ML-Leaks}{\cmark} &  & {\mlstab{Closed}} & \xmark & \cmark & $p$ & \xmark \\
            & & Li~\cite{li2019stealthy} & atk & {\scriptsize Evasion} & DL & \cmark & \cmark &  &  &  & \href{https://github.com/sli057/Video-Perturbation}{\cmark} &  &  & \cmark & \cmark & \xmark & $R$ \\
            & & Ma~\cite{ma2019nic} & def & {\scriptsize Evasion} & DL & \halfcirc & \cmark &  &  &  &  &  &  & \cmark & \cmark & $p$ & \xmark \\

            & & Li~\cite{li2018textbugger} & atk & {\scriptsize Evasion} & DL+SL & \halfcirc &  & \cmark &  &  &  & \cmark & {\mlstab{Closed}} & \xmark & \cmark & $p$ & $D$ \\

            \cline{2-18}
            
            & \multirow{3}{*}{\begin{tabular}{c} SP \\ {\tiny (3/84)}            \end{tabular} }
            & Ling~\cite{ling2019deepsec} & def & {\scriptsize Evasion} & DL & \cmark & \cmark &  &  &  & \href{https://github.com/ryderling/DEEPSEC}{\cmark} &  &  & \cmark & \cmark & $p$ & \xmark \\
            & & Wang~\cite{wang2019neural} & def & {\scriptsize Poison.} & DL & \cmark & \cmark &  &  &  & \href{https://github.com/bolunwang/backdoor}{\cmark} &  &  & \cmark & \cmark & \xmark & \cmark \\
            & & Nasr~\cite{nasr2019comprehensive} & atk & {\scriptsize Member.} & DL & \xmark & \cmark &  &  & {\scriptsize Finance} &  &  &  & \xmark & \cmark & $p$ & $D$ \\

            \cline{2-18}
            
            & \multirow{6}{*}{\begin{tabular}{c} SEC \\ {\tiny (6/113)}            \end{tabular} }
            & Tong~\cite{tong2019improving} & def & {\scriptsize Evasion} & DL+SL & \cmark &  &  &  & {\scriptsize Malware} & \href{https://github.com/mzweilin/PDF-Malware-Parser}{\cmark} &  &  & \xmark & \cmark & $p$ & \xmark \\
            & & Demontis~\cite{Demontis:Adversarial} & atk & {\scriptsize Evasion} & DL+SL & \xmark & \cmark &  &  & {\scriptsize Malware} &  &  &  & \xmark & \cmark & $p$ & $S$ \\
            & & Xiao~\cite{xiao2019seeing} & atk & {\scriptsize Evasion} & DL & \xmark & \cmark &  &  &  &  & \cmark & {\mlstab{Closed}} & \xmark & \cmark & $p$ & \xmark \\
            & & Quiring~\cite{quiring2019misleading} & atk & {\scriptsize Evasion} & DL+SL & \xmark &  & \cmark &  &  &  &  &  & \xmark & \cmark & $p$ & \xmark \\
            & & Hong~\cite{hong2019terminal} & atk & {\scriptsize Evasion} & DL & \xmark & \cmark &  &  &  &  & * &  & \xmark & \xmark & $p$ & \xmark \\
            & & Batina~\cite{batina2019csi} & atk & {\scriptsize Stealing} & DL & \halfcirc & \cmark &  &  &  &  & * &  & \xmark & \cmark & $p$ & \xmark \\

            \cline{2-18}
            
            & \multirow{10}{*}{\begin{tabular}{c} CCS \\ {\tiny (10/149)}  \end{tabular} }
            & Song~\cite{song2019privacy} & atk & {\scriptsize Member.} & DL & \xmark & \cmark &  &  &  & \href{https://github.com/inspire-group/privacy-vs-robustness}{\cmark} &  &  & \xmark & \cmark & $p$ & \xmark \\
            & & Jia~\cite{jia2019memguard} & def & {\scriptsize Member.} & DL & \xmark & \cmark &  &  & {+} &  &  &  & \xmark & \cmark & $p$ & \xmark \\
            & & Co~\cite{co2019procedural} & atk & {\scriptsize Evasion} & DL & \cmark & \cmark &  &  &  &  &  &  & \xmark & \cmark & $p$ & \xmark \\
            & & Liu~\cite{liu2019abs} & def & {\scriptsize Poison.} & DL & \cmark & \cmark &  &  &  &  &  &  & \cmark & \cmark & $p$ & \cmark \\
            & & Baluta~\cite{baluta2019quantitative} & def & {\scriptsize Poison.} & DL & \halfcirc & \cmark &  &  &  & \href{https://teobaluta.github.io/NPAQ/}{\cmark} &  &  & \cmark & \cmark & $p$ & \cmark \\
            & & Zhao~\cite{zhao2019seeing} & atk & {\scriptsize Evasion} & DL & \xmark & \cmark &  &  &  & \href{https://sites.google.com/view/ai-tricker}{\cmark} &  & {\mlstab{Open}} & \xmark & \cmark & $p$ & $S$ \\
            & & Tramer~\cite{tramer2019adversarial} & atk & {\scriptsize Evasion} & DL & \halfcirc & \cmark &  &  &  & \href{https://github.com/ftramer/ad-versarial}{\cmark} & \cmark & {\mlstab{Closed}} & \xmark & \cmark & $s$ & \xmark \\
            & & Wang~\cite{wang2019attacking} & atk & {\scriptsize Evasion} & DL & \cmark &  &  &  & {\scriptsize Graphs} &  &  &  & \xmark & \cmark & \xmark & $\subset$ \\
            & & Yao~\cite{yao2019latent} & atk & {\scriptsize Poison.} & DL & \halfcirc & \cmark &  &  &  &  &  &  & \halfcirc & \cmark & $p$ & $\subset$ \\
            & & Yang~\cite{yang2019neural} & atk & {\scriptsize Stealing} & DL & \halfcirc & \cmark &  &  &  &  &  & {\mlstab{Closed}} & \xmark & \cmark & $p$ & $D$ \\

            \midrule

            \multirow{24}{*}{\begin{tabular}{c} 2020 \\ {\tiny (24/470)}            \end{tabular}} & 
            \multirow{2}{*}{\begin{tabular}{c} NDSS \\ {\tiny (2/88)}            \end{tabular} }
            & Aghakhani~\cite{aghakhani2020malware} & atk & {\scriptsize Evasion*} & DL+SL & \xmark &  &  &  & {\scriptsize Malware} & \href{https://github.com/ucsb-seclab/packware}{\cmark} &  &  & \xmark & \cmark & \xmark & $D$ \\
            & & Yu~\cite{yu2020cloudleak} & atk & {\scriptsize Stealing} & DL & \cmark & \cmark &  &  &  &  &  & {\mlstab{Closed}} & \xmark & \cmark & $p$ & $D$ \\
            \cline{2-18}
            
            & \multirow{4}{*}{\begin{tabular}{c} SP \\ {\tiny (4/104)}            \end{tabular} }
            & Schuster~\cite{schuster2020humpty} & atk & {\scriptsize Poison.} & DL & \cmark &  & \cmark &  &  &  &  &  & \xmark & \cmark & \xmark & $\subset$ \\

            & & Pierazzi~\cite{Pierazzi:Intriguing} & atk & {\scriptsize Evasion} & SL & \cmark &  &  &  & {\scriptsize Malware} & \href{https://s2lab.cs.ucl.ac.uk/projects/intriguing/}{\cmark} & \cmark &  & \cmark & \cmark & $p$ & $R$ \\

            & & Chen~\cite{chen2020hopskipjumpattack} & def & {\scriptsize Evasion} & DL & \cmark & \cmark &  &  &  & \href{https://github.com/Jianbo-Lab/HSJA/}{\cmark} &  &  & \xmark & \cmark & $l$ & \xmark \\

            & & Jan~\cite{jan2020throwing} & atk & {\scriptsize Evasion*} & DL & \cmark &  &  &  & {\scriptsize Network} &  &  &  & \xmark & \cmark & $p$ & $\subset$ \\

            \cline{2-18}
            
            & \multirow{8}{*}{\begin{tabular}{c} SEC \\ {\tiny (8/157)}            \end{tabular} }
            & Salem~\cite{salem2020updates} & atk & {\scriptsize Member.} & DL & \xmark & \cmark &  &  & {\scriptsize Location} &  &  &  & \halfcirc & \cmark & $p$ & $S$ \\
            & & Chandrasekaran~\cite{chandrasekaran2020exploring} & atk & {\scriptsize Stealing} & DL+SL & \cmark & \cmark & \cmark &  & {+} &  &  &  & \halfcirc & \cmark & $p$ & \xmark \\
            & & Suya~\cite{suya2020hybrid} & atk & {\scriptsize Evasion} & DL & \cmark & \cmark &  &  &  & \href{https://github.com/suyeecav/Hybrid-Attack}{\cmark} &  &  & \xmark & \cmark & $p$ & $S$ \\
            & & Jagielski~\cite{jagielski2020high} & atk & {\scriptsize Stealing} & DL & \halfcirc & \cmark &  &  &  &  &  &  & \xmark & \cmark & $l$ & $D$ \\
            & & Quiring~\cite{quiring2020adversarial} & atk & {\scriptsize Evasion} & DL & \xmark & \cmark &  &  &  & \href{https://scaling-attacks.net/}{\cmark} & \cmark &  & \xmark & \cmark & $p$ & \xmark \\
            & & Li~\cite{li2020textshield} & def & {\scriptsize Evasion} & DL & \cmark &  & \cmark &  &  &  &  &  & \xmark & \cmark & $p$ & $D$ \\
            & & Leino~\cite{leino2020stolen} & atk & {\scriptsize Member.} & DL & \cmark & \cmark &  &  & {+} &  &  &  & \cmark & \cmark & $p$ & \xmark \\
            & & Zhang~\cite{zhang2020interpretable} & atk & {\scriptsize Evasion} & DL & \xmark & \cmark &  &  &  &  &  &  & \cmark & \cmark & $p$ & \xmark \\
 
            \cline{2-18}
            
            & \multirow{10}{*}{\begin{tabular}{c} CCS \\ {\tiny (10/121)}  \end{tabular} }
            & Nassi~\cite{nassi2020phantom} & atk & {\scriptsize Evasion} & DL & \xmark & \cmark &  &  &  & \href{https://github.com/ymirsky/GhostBusters}{\cmark} & \cmark & {\mlstab{Closed}} & \xmark & \cmark & $s$ & \xmark \\
            & & Li~\cite{li2020advpulse} & atk & {\scriptsize Evasion} & DL & \xmark &  &  & \cmark &  &  &  &  & \cmark & \cmark & $p$ & $\subset$ \\
            & & Shan~\cite{shan2020gotta} & def & {\scriptsize Evasion} & DL & \halfcirc & \cmark &  &  &  & \href{https://github.com/Shawn-Shan/trapdoor}{\cmark} &  &  & \cmark & \cmark & $p$ & \xmark \\
            & & Pang~\cite{pang2020tale} & atk & {\scriptsize Poison.} & DL & \cmark & \cmark &  &  &  & \href{https://github.com/alps-lab/imc}{\cmark} &  &  & \cmark & \cmark & $p$ & \cmark \\
            & & Abdelnabi~\cite{abdelnabi2020visualphishnet} & atk & {\scriptsize Evasion*} & DL & \xmark & \cmark &  &  & {\scriptsize Phishing} & \href{https://s-abdelnabi.github.io/VisualPhishNet/}{\cmark} & \cmark &  & \xmark & \cmark & \xmark & \xmark \\
            & & Li~\cite{li2020deepdyve} & atk & {\scriptsize Evasion} & DL & \cmark & \cmark &  &  &  &  &  &  & \cmark & \cmark & $p$ & \xmark \\
            & & Lin~\cite{lin2020composite} & atk & {\scriptsize Poison.} & DL & \cmark & \cmark & \cmark &  &  & \href{https://github.com/TemporaryAcc0unt/composite-attack}{\cmark} &  &  & \cmark & \cmark & \xmark & \cmark \\
            & & Chen~\cite{chen2020gan} & atk & {\scriptsize Member.} & DL & \cmark & \cmark &  &  &  &  &  &  & \xmark & \cmark & $p$ & $D$ \\
            & & Zanella~\cite{zanella2020analyzing} & atk & {\scriptsize Member.} & DL & \cmark &  & \cmark &  &  & \href{(broken link)}{\cmark} &  &  & \xmark & \cmark & $p$ & \xmark \\
            & & Song~\cite{song2020information} & atk & {\scriptsize Member.} & DL & \cmark &  & \cmark &  &  &  &  &  & \xmark & \cmark & $p$ & $S$ \\

            \midrule
            
            \multirow{40}{*}{\begin{tabular}{c} 2021 \\ {\tiny (41/644)}            \end{tabular}} & 
            \multirow{3}{*}{\begin{tabular}{c} NDSS \\ {\tiny (3/87)}            \end{tabular} }
            & Hui~\cite{hui2021practical} & atk & {\scriptsize Member.} & DL & \xmark & \cmark &  &  & {+} & \href{https://github.com/hyhmia/BlindMI}{\cmark} &  &  & \xmark & \cmark & $p$ & \xmark \\

            & & Huang~\cite{huang2021data} & atk & {\scriptsize Poison.} & DL & \cmark &  &  &  & {\scriptsize Ratings} &  &  &  & \xmark & \cmark & \xmark & $\subset$ \\

            & & Barradas~\cite{barradas2021flowlens} & atk & {\scriptsize Evasion*} & SL & \cmark &  &  &  & {\scriptsize Network} & \href{https://github.com/dmbb/FlowLens}{\cmark} &  &  & \xmark & \xmark & \xmark & \xmark \\

            \cline{2-18}
            
            & \multirow{5}{*}{\begin{tabular}{c} SP \\ {\tiny (5/115)}            \end{tabular} }
            & Xu~\cite{xu2021detecting} & def & {\scriptsize Poison.} & DL & \cmark & \cmark &  & \cmark &  & \href{https://github.com/AI-secure/Meta-Nerual-Trojan-Detection}{\cmark} &  &  & \cmark & \cmark & $p$ & \cmark \\

            & & Abdelnabi~\cite{abdelnabi2021adversarial} & atk & {\scriptsize Evasion} & DL & \halfcirc &  & \cmark &  &  & \href{https://github.com/S-Abdelnabi/awt/}{\cmark} & \cmark &  & \xmark & \cmark & $p$ & \xmark \\

            & & Chen~\cite{chen2021real} & atk & {\scriptsize Evasion} & DL & \cmark &  &  & \cmark &  & \href{https://sites.google.com/view/fakebob }{\cmark} & \cmark & {\scriptsize (both) } & \xmark & \cmark & $s$ & $S$ \\

            & & Abdullah~\cite{abdullah2021hear} & atk & {\scriptsize Evasion} & DL & \cmark &  &  & \cmark &  &  & \cmark & {\mlstab{Closed}} & \xmark & \cmark & \xmark & \xmark \\

            & & Nasr~\cite{nasr2021adversary} & def & {\scriptsize Evasion} & DL & \cmark & \cmark &  &  & {\scriptsize Finance} &  &  &  & \xmark & \cmark & $p$ & \xmark \\

            \cline{2-18}
            
            & \multirow{24}{*}{\begin{tabular}{c} SEC \\ {\tiny (24/246)}            \end{tabular} }
            & Sato~\cite{sato2021dirty} & atk & {\scriptsize Evasion} & DL & \cmark & \cmark &  &  &  & \href{https://sites.google.com/view/cav-sec/drp-attack}{\cmark} & \cmark & {\mlstab{Open}} & \cmark & \cmark & $p$ & \xmark \\
            & & Nasr~\cite{nasr2021defeating} & atk & {\scriptsize Evasion} & DL & \cmark &  &  &  & {\scriptsize Network} & \href{https://github.com/SPIN-UMass/BLANKET}{\cmark} & \cmark &  & \xmark & \cmark & $p$ & $D$ \\
            & & He~\cite{he2021stealing} & atk & {\scriptsize Member.} & DL & \halfcirc &  &  &  & {\scriptsize Graph} &  &  &  & \xmark & \cmark & $p$ & \xmark \\
            & & Severi~\cite{severi2021explanation} & atk & {\scriptsize Poison.} & DL+SL & \cmark &  &  &  & {\scriptsize Malware} & \href{https://github.com/ClonedOne/MalwareBackdoors}{\cmark} &  &  & \xmark & \cmark & $p$ & \cmark \\
            & & Bagdasaryan~\cite{bagdasaryan2021blind} & atk & {\scriptsize Poison.} & DL & \cmark & \cmark & \cmark &  &  & \href{https://github.com/ebagdasa/backdoors101}{\cmark} &  &  & \xmark & \xmark & \xmark & \xmark \\
            & & Xi~\cite{xi2021graph} & atk & {\scriptsize Poison.} & DL & \cmark &  &  &  & {\scriptsize Graph} & \href{https://github.com/ebagdasa/backdoors101}{\cmark} &  &  & \halfcirc & \cmark & \xmark & $S$ \\
            & & Tang~\cite{tang2021demon} & def & {\scriptsize Poison.} & DL & \xmark & \cmark &  &  &  & \href{https://github.com/TDteach/Demon-in-the-Variant}{\cmark} &  &  & \halfcirc & \cmark & $p$ & $\subset$ \\
            & & Schuster~\cite{schuster2021you} & atk & {\scriptsize Poison.} & DL & \cmark &  & \cmark &  &  &  &  & {\mlstab{Open}} & \xmark & \cmark & \xmark & $\subset$ \\
            & & Carlini~\cite{carlini2021poisoning} & atk & {\scriptsize Poison.} & DL & \cmark & \cmark &  &  &  &  &  &  & \xmark & \cmark & \xmark & $\subset$ \\
            & & Vicarte~\cite{vicarte2021double} & atk & {\scriptsize Poison.} & DL & \cmark & \cmark &  &  &  &  &  &  & \xmark & \cmark & $p$ & \xmark \\
            & & Lovisotto~\cite{lovisotto2021slap} & atk & {\scriptsize Evasion} & DL & \cmark & \cmark &  &  &  & \href{https://github.com/ssloxford/short-lived-adversarial-perturbations}{\cmark} &  & {\mlstab{Open}} & \xmark & \cmark & \xmark & $\subset$ \\
            & & Carlini~\cite{carlini2021extracting} & atk & {\scriptsize Member.} & DL & \cmark &  & \cmark &  &  &  & \cmark & {\mlstab{Open}} & \xmark & \cmark & $p$ & \xmark \\
            & & Han~\cite{han2021sigl} & atk & {\scriptsize Evasion*} & DL & \xmark &  &  &  & {\scriptsize Graph} &  &  &  & \xmark & \cmark & $p$ & \xmark \\
            & & Eisenhofer~\cite{eisenhofer2021dompteur} & def & {\scriptsize Evasion} & DL & \cmark &  &  & \cmark &  & \href{https://github.com/rub-syssec/dompteur}{\cmark} & \cmark & {\mlstab{Open}} & \cmark & \cmark & $p$ & $R$ \\
            & & Wu~\cite{wu2021adversarial} & atk & {\scriptsize Poison.} & DL & \cmark &  &  &  & {\scriptsize Games} & \href{https://github.com/psuwuxian/rl_attack}{\cmark} &  &  & \xmark & \cmark & $s$ & \xmark \\
            & & He~\cite{he2021drmi} & atk & {\scriptsize Stealing} & DL & \cmark & \cmark &  &  &  &  &  &  & \xmark & \cmark & $l$ & $S$ \\
            & & Rakin~\cite{rakin2021deep} & atk & {\scriptsize Evasion} & DL & \cmark & \cmark &  &  &  & \href{https://github.com/ASU-ESIC-FAN-Lab/DEEPDUPA}{\cmark} & * &  & \xmark & \cmark & $p$ & \xmark \\
            & & Jia~\cite{jia2021entangled} & def & {\scriptsize Stealing} & DL & \cmark & \cmark &  & \cmark &  & \href{https://github.com/cleverhans-lab/entangled-watermark}{\cmark} &  &  & \halfcirc & \cmark & $l$ & $\subset$ \\
            & & Zhu~\cite{zhu2021hermes} & def & {\scriptsize Stealing} & DL & \xmark & \cmark &  &  &  &  & * &  & \xmark & \cmark & $p$ & \xmark \\
            & & Xiang~\cite{xiang2021patchguard}& def & {\scriptsize Evasion} & DL & \cmark & \cmark &  &  &  & \href{https://github.com/inspire-group/PatchGuard}{\cmark} &  &  & \cmark & \cmark & $p$ & \xmark \\
            & & Lin~\cite{lin2021phishpedia} & atk & {\scriptsize Evasion *} & DL & \cmark & \cmark &  &  & {\scriptsize Phishing} & \href{https://sites.google.com/view/phishpedia-site/home}{\cmark} & \cmark &  & \xmark & \cmark & $p$ & \xmark \\
            & & Azizi~\cite{azizi2021t} & def & {\scriptsize Poison.} & DL & \xmark &  & \cmark &  &  & \href{https://github.com/reza321/T-Miner}{\cmark} &  &  & \cmark & \cmark & $p$ & \cmark \\
            & & Hussain~\cite{hussain2021waveguard} & def & {\scriptsize Evasion} & DL & \xmark &  &  & \cmark &  & \href{https://github.com/shehzeen/waveguard_defense}{\cmark} &  &  & \cmark & \cmark & $p$ & \xmark \\
            & & Song~\cite{song2021systematic}& def & {\scriptsize Member.} & DL & \cmark & \cmark &  &  & {+} & \href{https://github.com/inspire-group/membership-inference-evaluation}{\cmark} &  &  & \xmark & \cmark & $l$ & \xmark \\
            \cline{2-18}
            
            & \multirow{10}{*}{\begin{tabular}{c} CCS \\ {\tiny (9/196)}  \end{tabular} }
            & Zheng~\cite{zheng2021black} & atk & {\scriptsize Evasion} & DL & \halfcirc &  &  & \cmark &  &  & \cmark & {\mlstab{Closed}} & \xmark & \cmark & \xmark & \xmark \\
            & & Mu~\cite{mu2021hard} & atk & {\scriptsize Evasion} & DL & \cmark &  &  &  & {\scriptsize Graphs} &  &  &  & \xmark & \cmark & $l$ & \xmark \\
            & & Bahramali~\cite{bahramali2021robust} & atk & {\scriptsize Evasion} & DL & \cmark &  &  &  & {\scriptsize Network} &  &  &  & \xmark & \cmark & \xmark & $S$ \\
            & & Sheatsley~\cite{sheatsley2021robustness} & atk & {\scriptsize Evasion} & DL & \xmark &  &  &  & {\scriptsize Network} &  &  &  & \cmark & \cmark & $p$ & $R$ \\
            & & Du~\cite{du2021cert} & def & {\scriptsize Evasion} & DL & \cmark & \cmark & \cmark &  &  &  &  &  & \cmark & \cmark & $p$ & $R$ \\
            & & Li~\cite{li2021tss} & def & {\scriptsize Evasion} & DL & \cmark & \cmark &  &  &  & \href{https://github.com/AI-secure/semantic-randomized-smoothing}{\cmark} &  &  & \xmark & \cmark & \xmark & \xmark \\
            & & He~\cite{he2021quantifying} & def & {\scriptsize Member.} & DL & \xmark & \cmark &  &  &  & \href{https://github.com/xinleihe/ContrastiveLeaks}{\cmark} &  &  & \halfcirc & \cmark & $l$ & $S$ \\
            & & Li~\cite{li2021membership} & atk & {\scriptsize Member.} & DL & \cmark & \cmark &  &  &  & \href{https://github.com/zhenglisec/Decision-based-MIA}{\cmark} &  &  & \xmark & \cmark & $l$ & \xmark \\
            & & Chen~\cite{chen2021machine} & def & {\scriptsize Member.} & DL+SL & \xmark & \cmark &  &  & {+} & \href{https://github.com/MinChen00/UnlearningLeaks}{\cmark} &  &  & \xmark & \cmark & $p$ & \xmark \\

            \midrule
            
            \multirow{2}{*}{\begin{tabular}{c} \textbf{Year} \\ (subs)  \end{tabular}} &
            \multirow{2}{*}{\begin{tabular}{c} \textbf{Venue} \\ (subs)  \end{tabular}} &
            \multirow{2}{*}{\begin{tabular}{c} \textbf{Paper} \\ (1st author)  \end{tabular}} & 
            
            \textit{Focus} & \textit{Attack} & \textit{Paradigm} & \textit{Cost} & \textit{Img} & \textit{Text} & \textit{Audio} & \textit{Other} & \textit{Code} & \textit{Pipeline} & \textit{Type} & \textit{Param.} & \textit{Sem.} & \textit{Output} & \textit{Training} \\ \cline{4-18}
            
            & & &
            \multicolumn{1}{c|}{\textbf{G1}} & 
            \multicolumn{1}{c|}{\textbf{G2}} &
            \multicolumn{1}{c|}{\textbf{G3}} & 
            \multicolumn{1}{c|}{\textbf{G4}} &
            \multicolumn{4}{c|}{\textbf{G5} (Evaluation Data)} &
            \multicolumn{1}{c|}{\textbf{G6}} & 
            \multicolumn{1}{c|}{\textbf{G7}} & 
            \multicolumn{1}{c?}{\textbf{G8}} &
            \multicolumn{1}{c|}{\textbf{T1}} &
            \multicolumn{1}{c|}{\textbf{T2}} &
            \multicolumn{1}{c|}{\textbf{T3}} &
            \multicolumn{1}{c}{\textbf{T4}} \\

            \bottomrule
        \end{tabular}
    }
\end{table*}

\end{document}